\RequirePackage[2020-02-02]{latexrelease}

\documentclass[twocolumn,preprintnumbers]{revtex4}%
\usepackage{amssymb}
\usepackage{amsmath}
\usepackage{graphicx}
\usepackage{epstopdf}
\usepackage{dcolumn}
\usepackage{bm}
\usepackage{xcolor}
\usepackage{ifthen}
\usepackage[normalem]{ulem}
\usepackage{amsfonts}%
\RequirePackage[2020-02-02]{latexrelease}
\graphicspath{{c:/Users/Eyal/MyDocuments/latex/Texfiles/Diamond/dipolarBA/EntangSupp/SDIP/}
	{c:/Users/Eyal/MyDocuments/latex/Texfiles/Diamond/dipolarBA/EntangSupp/SDH/}
	{c:/Users/Eyal/MyDocuments/latex/Texfiles/Diamond/fMixYIG/}
	{c:/Users/Eyal/MyDocuments/latex/Texfiles/Diamond/FerriMO/YIGS/}}
\UseRawInputEncoding
\providecommand{\U}[1]{\protect\rule{.1in}{.1in}}
\providecommand{\U}[1]{\protect\rule{.1in}{.1in}}
\def\showal{1}
\newcommand{\al}[1]{\ifthenelse{\showal=1}{\textcolor{orange}{[[#1]]}}{}}

\newcommand{\eb}[1]{\ifthenelse{\showal=1}{\textcolor{cyan}{[[#1]]}}{}}
\begin{document}
\title{Disentanglement--induced bistability in a magnetic resonator}
\author{Eyal Buks}
\email[]{eyal@ee.technion.ac.il}
\affiliation{Andrew and Erna Viterbi Department of Electrical Engineering, Technion, Haifa
32000 Israel}
\date{\today }

\begin{abstract}
Multi--stability in the response of a ferrimagnetic spin resonator to an externally applied driving is experimentally studied. The observed multi--stability cannot be derived from any master equation that linearly depends on the spins' reduced density operator. Traditionally, the nonlinearity that is required in order to theoretically account for the observed multi--stability is introduced by implementing the method of Bosonization. Here, an alternative explanation, which is based on the hypothesis that disentanglement spontaneously occurs in quantum systems is explored. According to this hypothesis, time evolution is governed by a master equation having an added nonlinear term, which deterministically generates disentanglement. Experimental results are compared with predictions derived from both competing theoretical models. It is found that better agreement with data is obtained from the disentanglement--based model. This finding, together with a difficulty to justify the Bosonization--based model, indirectly support the spontaneous disentanglement hypothesis.
\end{abstract}
\maketitle

\textbf{Introduction} -- A variety of nonlinear extensions to standard quantum
mechanics (QM) have been proposed
\cite{Weinberg_61,Doebner_3764,Gisin_5677,Gisin_2259,Kaplan_055002,Munoz_110503,Jacobs_279,Geller_2200156}%
. These proposals are mainly motivated by an apparent internal inconsistency
in QM, which was first introduced in 1935 by Schr\"{o}dinger
\cite{Schrodinger_807}, and which is commonly known as the problem of quantum
measurement
\cite{Penrose_4864,Bassi_471,Pearle_857,Ghirardi_470,Bassi_257,Bennett_170502,Kowalski_1,Fernengel_385701,Kowalski_167955,Oppenheim_041040,Schrinski_133604}%
.

Some of the proposed nonlinear extensions to QM can be experimentally tested
by studying the stability of quantum systems. Time evolution is commonly
derived in standard QM from a master equation that linearly depends on the
reduced density operator $\rho$ \cite{Lindblad_119}. This linear dependency
excludes multistabilities in finite quantum systems \cite{Buks_012439}. In
contrast, multistabilities are experimentally observed in a variety of quantum
systems, including molecular--size ones
\cite{Roch_633,Thomas_145,Trishin_236801,Blesio_045113,Yamasaki_1187,Venkataramani_445}%
. Moreover, experimental observations of processes such as phase transitions
\cite{Chomaz_68,mainwood2005phase,Callender_539,Liu_S92,Ardourel_99,Shech_1170,toda1978statistical_I,Sakthivadivel_035,Vojta_2069}
and dynamical instabilities \cite{Suhl_209}, are arguably inconsistent with
linear dynamics.

Here, a recently--proposed nonlinear extension, which
is based on the hypothesis that disentanglement spontaneously occurs in
quantum systems [see the modified master equation (\ref{MME}) below]
\cite{Buks_2400036}, is experimentally explored. This hypothesis makes the collapse postulate of standard
QM redundant. Moreover, it can account for multistability in both static and
dynamical quantum systems \cite{Buks_012439}. The hypothesis is falsifiable,
since Eq. (\ref{MME}) yields predictions that are distinguishable from what is
derived from standard QM. In the current study the spontaneous disentanglement hypothesis is experimentally tested using a ferrimagnetic sphere resonator (FMSR) \cite{Stancil_Spin}. This magnetically--tunable spin system has a
variety of applications in the fields of magnonics
\cite{Suhl_209,Zheng_151101,Rameshti_1,Kusminskiy_299,Wang_057202,Wang_224410,Hyde_174423,Juraschek_094407,Lima_1}%
, opto-magnonics
\cite{Kusminskiy_1911_11104,Sharma_087205,Zhu_2012_11119,Bittencourt_014409},
and quantum data processing
\cite{Lachance_070101,Lachance_1910_09096,Tabuchi_729,Elyasi_054402,Zhang_023021}%
.

\begin{figure}[ptb]
\begin{center}
\includegraphics[width=3.0in,keepaspectratio]{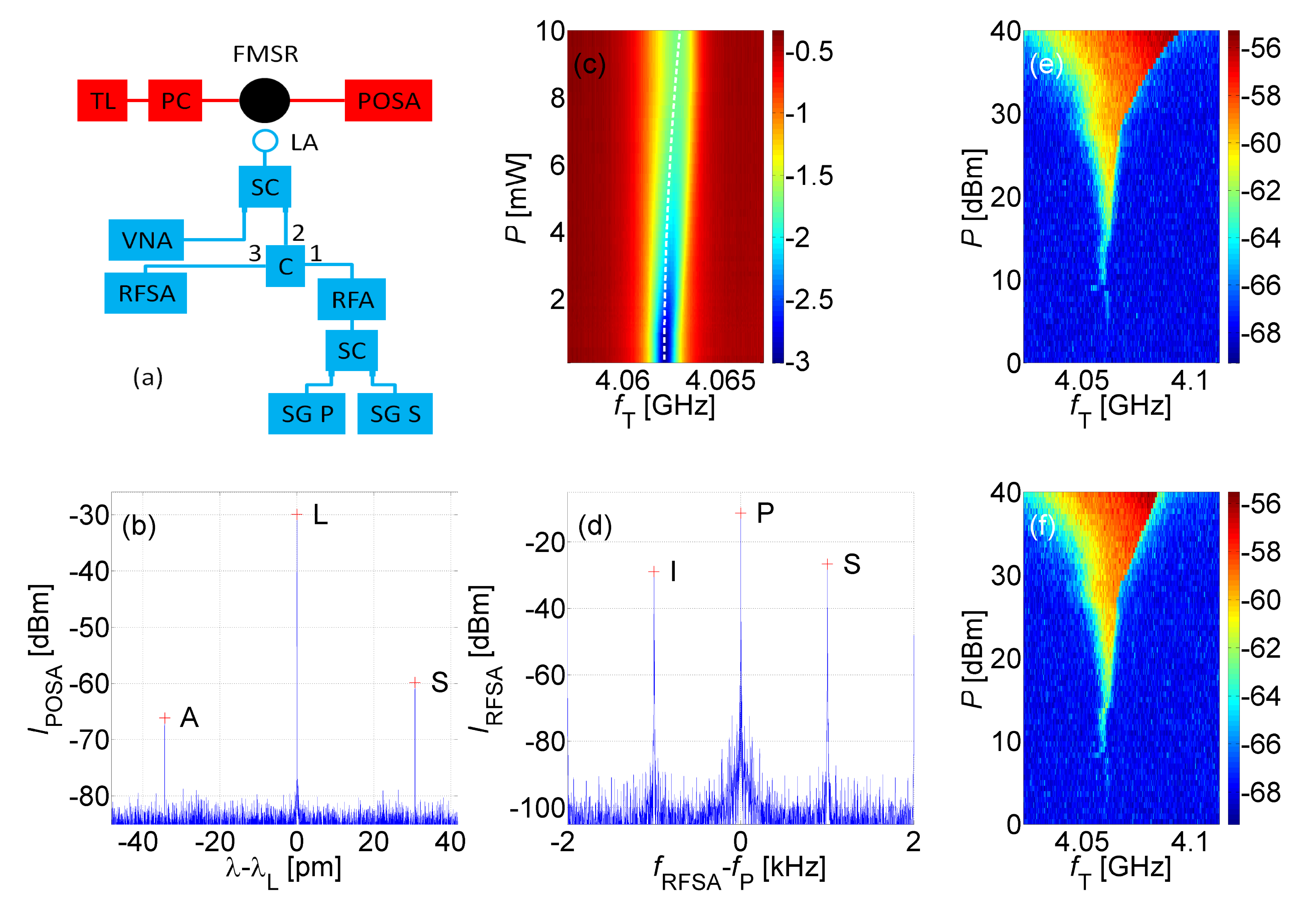}
\end{center}
\caption{{}Experimental setup. (a) RF components and coaxial cables are blue
colored, whereas the color red is used to label optical components and single
mode optical fibers. The externally--applied static magnetic field
$\mathbf{H}_{\mathrm{s}}$ (not shown in the sketch) is normal to the unit
vector $\mathbf{\hat{k}}$ pointing in the light propagation direction, and the
LA RF driving magnetic field is nearly parallel to $\mathbf{\hat{k}}$. A FMSR
made of YIG having radius of $R_{\mathrm{s}}=125\operatorname{   \mu m}$ is
held by two ceramic ferrules (not shown in the sketch), which provide
transverse alignment for both input and output single mode optical fibers. All
optical measurements are performed in the telecom band, in which YIG has
refractive index of $2.19$ and absorption coefficient of $\left(  0.5
\operatorname{m} \right)  ^{-1}$ \cite{Onbasli_1}. The state of polarization
of light illuminating the FMSR is controlled by a PC. All measurements are
performed at room temperature. (b) POSA measurement of MO modulation. The
laser, Stokes and anti-Stokes peaks are labeled by the letters L, S and A,
respectively. The corresponding optical wavelengths are $\lambda_{\mathrm{L}}%
$, $\lambda_{\mathrm{L}}\left(  1+\lambda_{\mathrm{L}}f_{\mathrm{T}%
}/c)\right)  $ and $\lambda_{\mathrm{L}}\left(  1-\lambda_{\mathrm{L}%
}f_{\mathrm{T}}/c)\right)  $, respectively, where $\lambda_{\mathrm{L}}=1537.7
\operatorname{nm}$ is the laser wavelength, $f_{\mathrm{T}}$ is the RF driving
frequency, and $c$ is the vacuum speed of light. Optical power illuminating
the FMSR is $10\operatorname{mW}$, and RF power injected into the LA is $18
\operatorname{dBm}$ (c) FMSR reflectivity $\left|  r\right|  ^{2}$ in dB units
measured using a VNA. (d) RFSA measurement of IMD. The pump, signal and idler peaks are
labeled by the letters P, S and I, respectively [the pump and signal tones are
generated by SG P and SG S, respectively, see (a)]. (e) and (f)
Bistability--induced hysteresis in the MO modulation S peak intensity [see
(b)] as a function of the driving frequency $f_{\mathrm{T}}$ and driving power
$P$. For (e) and (f), the driving frequency $f_{\mathrm{T}}$ is swept upwards
and downwards, respectively, and S peak intensity is plotted in dBm units.}%
\label{FigSetup}%
\end{figure}

Bistability in the FMSR response to externally--applied transverse driving
[see Fig. \ref{FigSetup}(e-f) below] has been studied in \cite{Wang_224410}.
This experimentally--observed bistability cannot be derived from any master
equation that linearly depends on the reduced density operator $\rho$. All
proposed theoretical explanations for multistabilities in finite quantum systems
are based on the assumption that time evolution is nonlinear. For spin
systems, nonlinearity can be introduced by implementing a method called
Bosonization, which is based on the Holstein--Primakoff transformation
\cite{Holstein_1098}. For driven spins, this method yields a theoretical
description that is analogous to the Duffing--Kerr model. The nonlinear term
in this model originates from magnetic anisotropy \cite{Wang_224410}.

The current study is mainly motivated by a difficulty to justify the Bosonization--based model, which enables multistabilities that are otherwise excluded. Here, an alternative theoretical model, which is based on the spontaneous
disentanglement hypothesis, is proposed and explored. Both
competing theoretical models yield cubic polynomial equations for the system's
steady state [see Eqs. (\ref{RDM cubic}) and (\ref{DKM E}) below]. The region
where bistability is theoretically expected is mapped using stability analysis
applied to both cubic equations (see Figs. \ref{FigRD} and \ref{FigDK} below).
Predictions derived from the two competing theoretical models are compared
with the FMSR experimentally--observed response.

\textbf{Spontaneous disentanglement} -- The modified master equation for the
time evolution of the reduced density operator $\rho$ is given by
\cite{Grimaudo_033835,Buks_2400036,Kowalski_167955,Elben_200501,Sergi_1350163,Brody_230405}%
\begin{equation}
\frac{\mathrm{d}\rho}{\mathrm{d}t}=i\hbar^{-1}\left[  \rho,\mathcal{H}\right]
+\mathcal{L}-\Theta\rho-\rho\Theta+2\left\langle \Theta\right\rangle \rho\;,
\label{MME}%
\end{equation}
where $\hbar$ is the Planck's constant, $\mathcal{H}^{{}}=\mathcal{H}^{\dag}$
is the Hamiltonian, $\mathcal{L}$ is a Lindblad superoperator
\cite{Lindblad_119}, $\Theta^{{}}=\Theta^{\dag}$ and $\left\langle
\Theta\right\rangle =\operatorname{Tr}\left(  \Theta\rho\right)  $. The added
term $-\Theta\rho-\rho\Theta+2\left\langle \Theta\right\rangle \rho$ in Eq.
(\ref{MME}), where $\Theta$ is a $\rho$--dependent disentanglement operator,
gives rise to nonlinear dynamics. The construction of both the Lindblad
superoperator $\mathcal{L}$ and the disentanglement operator $\Theta$ is
explained in section \ref{AppMME} of the supporting information (SI) (see also
Ref. \cite{Buks_2400036}). The disentanglement process is characterized by a
rate denoted by $\gamma_{\mathrm{D}}$. The coupling between the spins and
their environment, which is accounted for by the Lindblad superoperator
$\mathcal{L}$, is characterized by energy--relaxation $\Gamma_{1}$ and
dephasing $\Gamma_{\varphi}$ rates, thermal occupation factor $\hat{n}_{0}$,
and longitudinal $T_{1}$ and transverse $T_{2}$ relaxation times [see SI section
\ref{AppMME}].

\textbf{Driven $L$ spin system} -- The current study explores the effect of
disentanglement on a system composed of $L$ coupled spins 1/2. The total
angular momentum vector operator $\mathbf{S}=\left(  S_{x},S_{y},S_{z}\right)
$ in units of $\hbar/2$ is given by $\mathbf{S}=\sum_{l=1}^{L}\mathbf{S}_{l}%
$, where $\mathbf{S}_{l}=\left(  S_{l,x},S_{l,y},S_{l,z}\right)  $ is the
$l$'th spin angular momentum vector operator. The closed-system Hamiltonian
$\mathcal{H}$ is assumed to be given by%
\begin{align}
\frac{\mathcal{H}}{\hbar}  &  =-\frac{\omega_{0}S_{z}}{2}\nonumber\\
&  +\frac{\omega_{\mathrm{K}}\left(  S_{+}S_{-}+S_{-}S_{+}\right)
+\omega_{\mathrm{A}}\left(  S_{+}^{2}+S_{-}^{2}\right)  }{8}\nonumber\\
&  +\frac{\omega_{1}\left(  S_{+}e^{i\omega_{\mathrm{T}}t}+S_{-}
e^{-i\omega_{\mathrm{T}}t}\right)  }{4}\;,\nonumber\\
&  \label{H DLS}%
\end{align}
where the rates $\omega_{0}$, $\omega_{\mathrm{K}}$, $\omega_{\mathrm{A}}$,
$\omega_{1}$ and $\omega_{\mathrm{T}}$\ are real constants, and where $S_{\pm
}=S_{x}\pm iS_{y}$. The terms $S_{+}S_{-}+S_{-}S_{+}=2\left(  S_{x}^{2}%
+S_{y}^{2}\right)  $ and $S_{+}^{2}+S_{-}^{2}=2\left(  S_{x}^{2}-S_{y}%
^{2}\right)  $ account for magnetic anisotropy. In the rotating wave
approximation the term proportional to $\omega_{\mathrm{A}}$ is disregarded.
For the case $\omega_{\mathrm{A}}=0$, the term proportional to $\omega
_{\mathrm{K}}$ gives rise to easy axis (plane) for $\omega_{\mathrm{K}}>0$
($\omega_{\mathrm{K}}<0$). Heisenberg equations of motion that are derived
from the Hamiltonian (\ref{H DLS}) are given in SI section \ref{AppDLS}.

\begin{figure}[ptb]
\begin{center}
\includegraphics[width=3.0in,keepaspectratio]{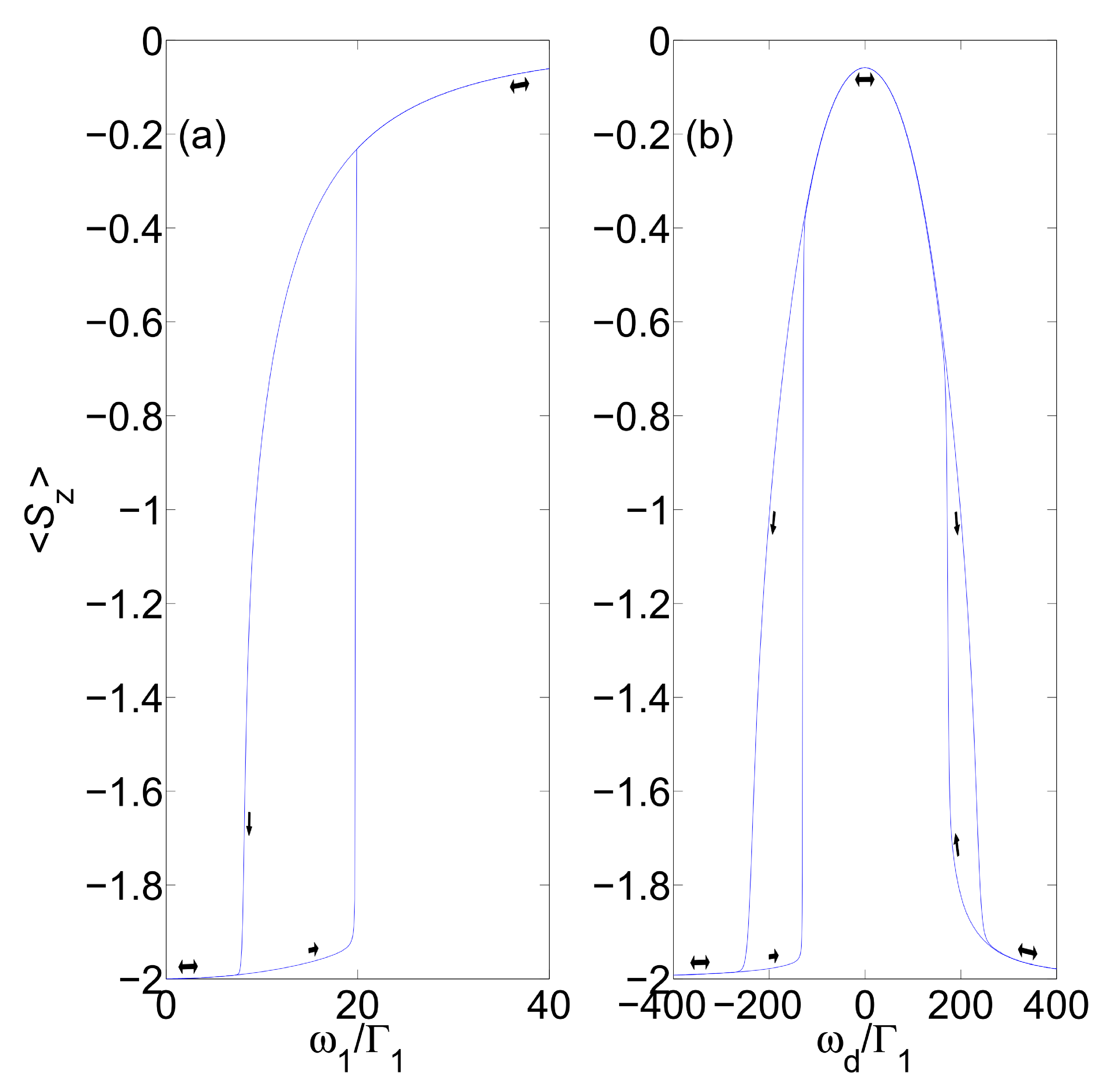}
\end{center}
\caption{{}Driven two spins. The expectation value $\left\langle
S_{z}\right\rangle $ in steady state plotted as a function of (a) $\omega
_{1}/\Gamma_{1}$ and (b) $\omega_{\mathrm{d}}/\Gamma_{1}$ (the angular
detuning frequency $\omega_{\mathrm{d}}$ is defined by $\omega_{\mathrm{d}
}=\omega_{\mathrm{T}}-\omega_{0}$). The overlaid arrows indicate the sweep
direction. Assumed parameters' values are $\omega_{\mathrm{K}}/\Gamma_{1}
=100$, $\omega_{\mathrm{A}}=0$, $\gamma_{\mathrm{D}}/\Gamma_{1}=100$,
$\Gamma_{\varphi}/\Gamma_{1}=0.1$, $\hat{n}_{0}=10^{-4}$, and $\omega
_{\mathrm{d}}/\Gamma_{1}=10$ for (a), and $\omega_{1}/\Gamma_{1}=40$ for (b).}%
\label{FigDTS}%
\end{figure}

In a frame rotating at the angular driving frequency $\omega_{\mathrm{T}}$,
the Hamiltonian (\ref{H DLS}) becomes time independent. Note that the Lindblad
superoperator $\mathcal{L}$ linearly depends on $\rho$ (see SI section
\ref{AppMME}). Thus, in the absence of disentanglement, i.e. for
$\Theta=0$, the master equation (\ref{MME}) has a unique steady state
solution. More generally, for any quantum system having Hilbert space of
finite dimensionality and time--independent Hamiltonian, multistability is
theoretically excluded, unless the master equation has nonlinear dependency on
$\rho$ \cite{Buks_012439}. In contrast, bistability is experimentally observed
in a variety of spin systems \cite{Cacchiani_5695,Aubay_15023}. On the other
hand, as is shown below, the disentanglement nonlinear term added in the
master equation (\ref{MME}) can give rise to multistability.

Disentanglement--induced bistability and hysteresis, which are excluded by
standard QM, are demonstrated by the plots shown in Fig. \ref{FigDTS}, which
display steady state solutions of the modified master equation (\ref{MME}) for
the case $L=2$ (i.e. two spins). Assumed parameters' values are listed in the
caption of Fig. \ref{FigDTS}. Matrix representation of the Hamiltonian
(\ref{H DLS}) for this case is given in SI section \ref{AppDTS}. The
plots shown in Fig. \ref{FigDTS} demonstrate that disentanglement can account
for bistability for the case $L=2$, however, for the FMSR under study here
$L\gg1$.

\textbf{Rapid disentanglement model} -- Quantum models of interacting spins
are commonly intractable, unless the number of spins $L$ is kept sufficiently
small. The added nonlinear term in the modified master equation (\ref{MME})
further complicates the dynamics. However, on the other hand, when the
disentanglement rate $\gamma_{\mathrm{D}}$ is sufficiently large, dynamics can
be significantly simplified by employing the rapid disentanglement (RD)
approximation, which is discussed in SI section \ref{AppRDM}. In this
approximation, the expectation value $P_{z}=\left\langle S_{z}\right\rangle $
satisfies in steady state a cubic polynomial equation given by [see Eq.
(\ref{eq for P_z}) in SI section \ref{AppRDM}]%
\begin{equation}
\frac{P_{z}}{P_{z0}}=\frac{1+\left(  \omega_{\mathrm{d}}-\omega_{\mathrm{K}
}P_{z}\right)  ^{2}T_{2}^{2}}{1+\left(  \omega_{\mathrm{d}}-\omega
_{\mathrm{K}}P_{z}\right)  ^{2}T_{2}^{2}+\omega_{1}^{2}T_{1}T_{2}}\;,
\label{RDM cubic}%
\end{equation}
where $P_{z0}$ represents the steady state value of $P_{z}$ for the case where
$\omega_{1}=0$ (no driving), and $\omega_{\mathrm{d}}=\omega_{\mathrm{T}%
}-\omega_{0}$ is the driving detuning angular frequency.

Bistability occurs in the region where the cubic polynomial equation
(\ref{RDM cubic}) has three real solutions (two of which representing locally
stable steady states). In the plane of normalized driving detuning and power,
this bistability region is red--colored in the stability map shown in Fig.
\ref{FigRD}(a). The normalized driving detuning is given by $\delta
=\omega_{\mathrm{d}}T_{2}$, the normalized driving power by $W=\omega_{1}%
^{2}T_{1}T_{2}/2$, and the normalized spin polarization by $z=P_{z}/P_{z0}$.
The boundary of bistability region contains two cusp (bistability onset)
points $\left(  \delta_{\pm},W\pm\right)  $, labeled in Fig. \ref{FigRD}(a) by
the symbols $P_{\pm}$. Analytical expressions for $\delta_{\pm}$ and $W\pm$
are derived in SI section \ref{AppRDM}. Bistability can occur provided
that $D\equiv\left(  \omega_{\mathrm{K}}T_{2}P_{z0}/4\right)  ^{2}>1$. The
plot in Fig. \ref{FigRD}(c) displays $\delta$--$z$ (detuning--polarization)
curves for five different values of $W$ (driving power), which are labeled by
the overlaid horizontal white dashed lines in Fig. \ref{FigRD}(a). For each
$\delta$ - $z$ curve, the point at which $\mathrm{d}z/\mathrm{d}\delta=0$ is
referred to as a peak point. The normalized detuning at the peak point, which
is denoted by $\delta_{\mathrm{p}}$, is plotted in Fig. \ref{FigRD}(b).

In the classical limit, the Hamiltonian (\ref{H DLS}) yields Poisson equations
of motion, which are analyzed in SI section \ref{AppNut}. The plot
shown in \ref{FigRD}(d) demonstrates a limit cycle steady state solution of
the classical equations of motion. In this plot, the overlaid color on the
surface of the polarization unit sphere represents the term $-\left(
\omega_{0}/2\right)  S_{z}+\left(  \omega_{\mathrm{K}}/8\right)  \left(
S_{+}S_{-}+S_{-}S_{+}\right)  +\left(  \omega_{\mathrm{A}}/8\right)  \left(
S_{+}^{2}+S_{-}^{2}\right)  $ in the Hamiltonian (\ref{H DLS}). The term
proportional to $\omega_{\mathrm{A}}$ gives rise to nutation and wobbling
(non--circular shape of the limit cycle).

\begin{figure}[ptb]
\begin{center}
\includegraphics[width=3.0in,keepaspectratio]{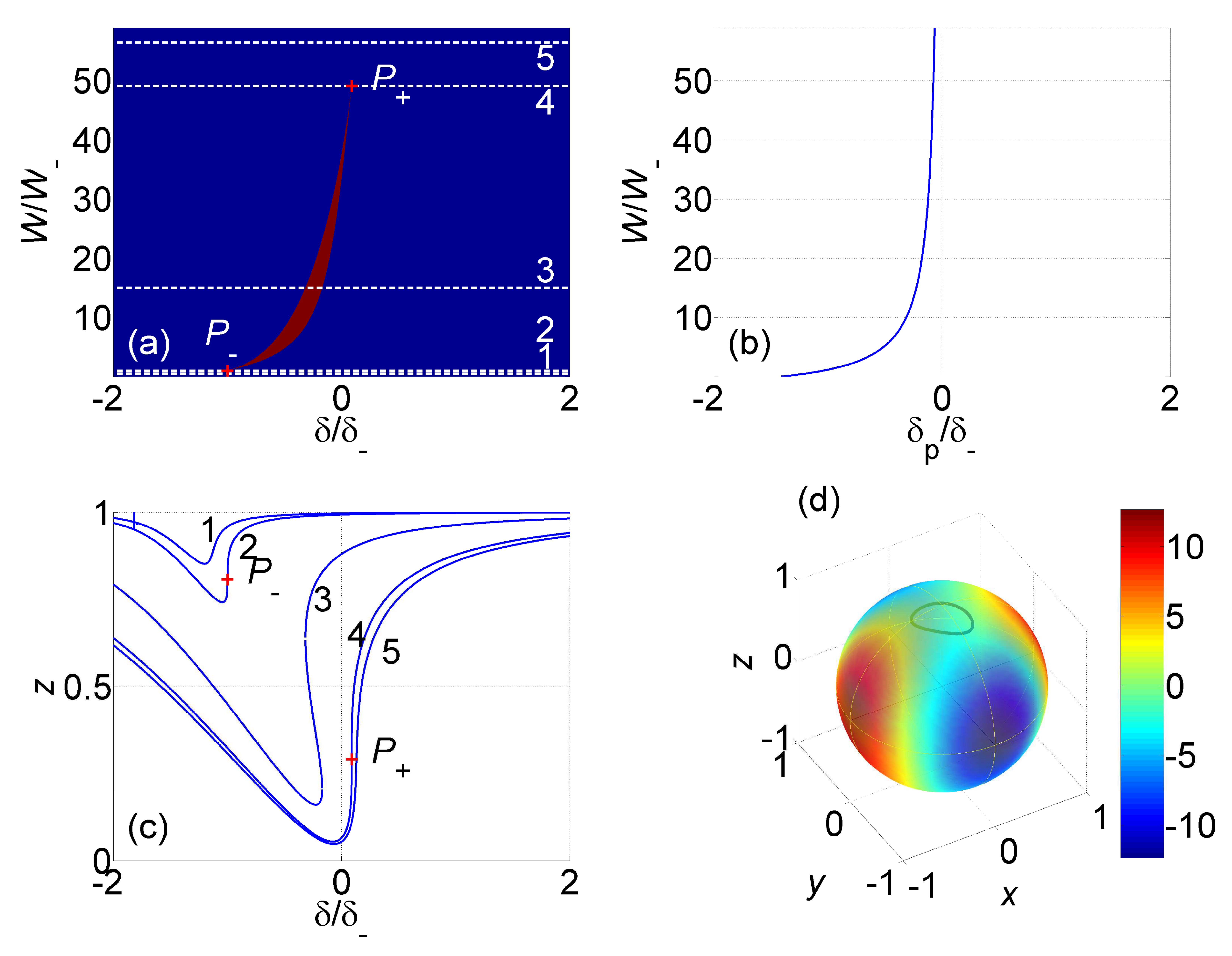}
\end{center}
\caption{{}RD model. (a) Stability map in the plane of normalized driving
detuning $\delta$ and power $W$. Calculation of steady state is based
on Eq. (\ref{RDM cubic}). Bistability occurs in the red-colored region. For
(a-c) $D=3$. (b) The normalized detuning at the peak point $\delta
_{\mathrm{p}}$ [see Eq. (\ref{RDM delta PP}) in SI section \ref{AppRDM}]. (c) Normalized spin polarization $z$ as a function of normalized detuning
$\delta$ for five different values of $W$, which are labeled by the overlaid
horizontal white dashed lines in (a). (d) Spin nutation in the classical
limit. The black curve represents a steady state limit cycle. For this example
$\omega_{\mathrm{d}}/\omega_{0}=10^{-5}$, $\omega_{1}/\omega_{0}=30$,
$\omega_{\mathrm{K}}/\omega_{0}=0.5$, $\omega_{\mathrm{A}}/\omega_{0}=50$,
$\omega_{0}T_{1}=\omega_{0}T_{2}=10^{-2}$, and $P_{z0}=0.9$.}%
\label{FigRD}%
\end{figure}

\textbf{Duffing--Kerr model} -- As was pointed out above, the linearity of
standard QM excludes multistability in finite quantum systems. On the other
hand, some commonly--employed approximation methods can give rise to nonlinear
dynamics. The method of Bosonization, which is based on the
Holstein--Primakoff transformation \cite{Holstein_1098}, maps the spins'
Hilbert space having finite dimensionality into a space having infinite
dimensionality. This method introduces nonlinearity that can give rise to
bistability in the presence of magnetic anisotropy \cite{Wang_224410}.
Justification of this method, which yields bistability that is otherwise
excluded, is arguably questionable.

\begin{figure}[ptb]
\begin{center}
\includegraphics[width=3.0in,keepaspectratio]{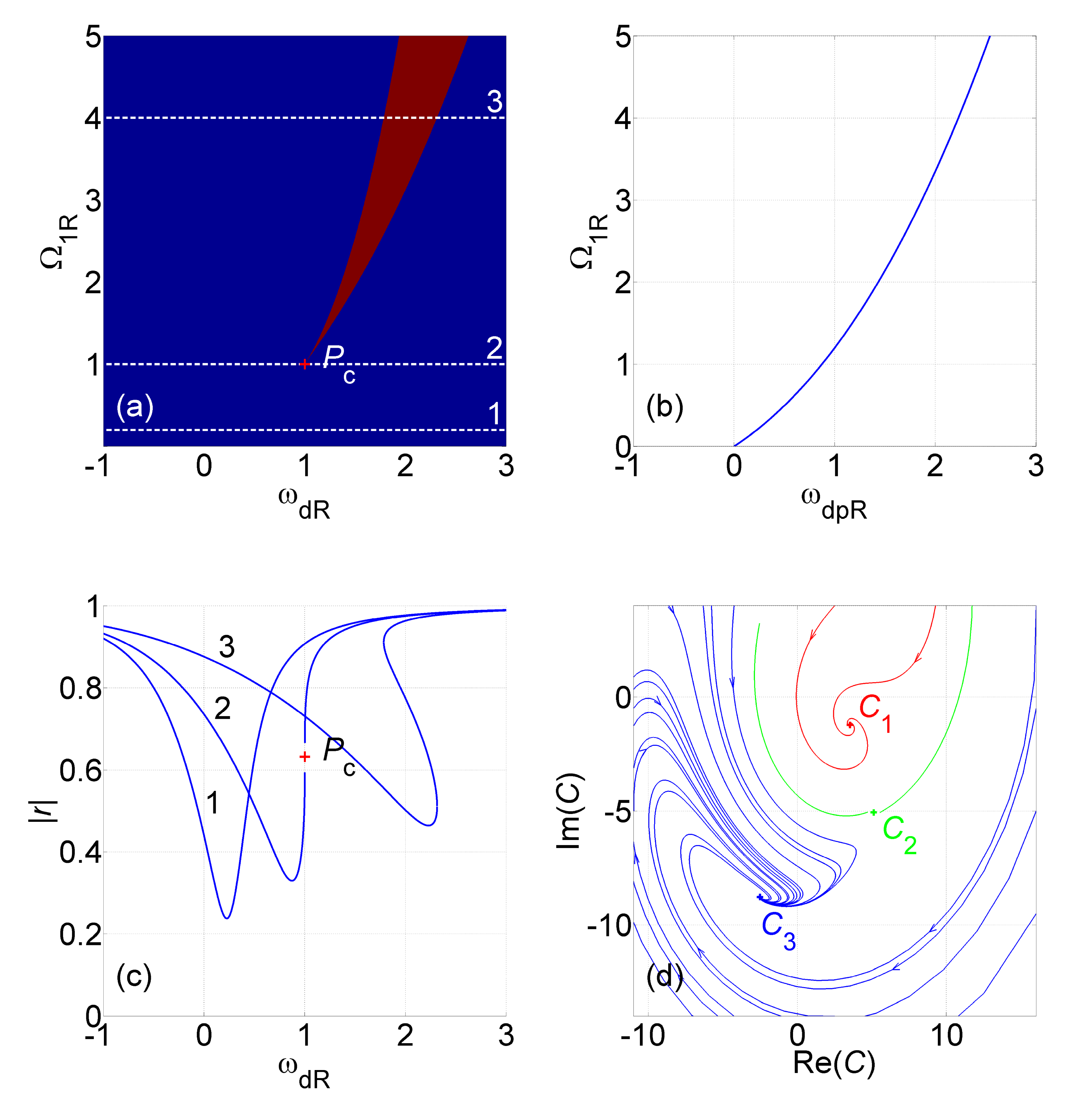}
\end{center}
\caption{{}Duffing-Kerr model. (a) Stability map in the plane of normalized
driving detuning $\omega_{\mathrm{dR}}=\omega_{\mathrm{d}}/\omega
_{\mathrm{dc}}$ and power $\Omega_{\mathrm{1R}}=\Omega_{1}/\Omega
_{1\mathrm{c}}$ [see Eqs. (\ref{omdc}) and (\ref{om1c}) in SI section
\ref{AppDKM}]. Bistability occurs in the red-colored region. The
bistability onset point is labeled by the symbol $P_{\mathrm{c}}$. (b) The
normalized detuning at the peak point $\omega_{\mathrm{dpR}}=\omega
_{\mathrm{dp}} /\omega_{\mathrm{dc}}$ ($\omega_{\mathrm{dp}}$ is the value of
$\omega_{\mathrm{d}}$ at a peak point). (c) Reflectivity $\left\vert
r\right\vert $ as a function of normalized detuning $\omega_{\mathrm{dR}}$
[see Eq. (\ref{DKM r}) in SI section \ref{AppDKM}]. For (a-c) assumed
parameters' values are $\gamma_{1}/\gamma=0.4$, $\gamma_{3}/\gamma=0.01$ and
$\omega_{\mathrm{K} }/\gamma=-0.1$. (d) Flow map of the Bosonic mode complex
amplitude $C$\ in the bistability region. The two locally stable steady states
are labelled as $C_{1}$ and $C_{3}$, and the unstable one (saddle point) as
$C_{2}$. The red and blue lines represent flow toward the spiral attractors at
$C_{1}$ and $C_{3}$, respectively. The green line is the seperatrix, namely
the boundary between the basins of attraction of the attractors at $C_{1}$ and
$C_{3}$.}%
\label{FigDK}%
\end{figure}

The Bosonization method \cite{Wang_224410}, which is reviewed in SI section
\ref{AppDKM}, yields in steady state a cubic polynomial equation for
$E=\left\vert C\right\vert ^{2}$ given by [see Eq. (\ref{eq for E}) in SI section
\ref{AppDKM}]%
\begin{equation}
E=\frac{2\gamma_{1}\Omega_{1}}{\left(  \omega_{\mathrm{d}}-\omega_{\mathrm{K}%
}E\right)  ^{2}+\left(  \gamma+\gamma_{3}E\right)  ^{2}}\;,\label{DKM E}%
\end{equation}
where $C$ is the Bosonic mode complex amplitude. Damping is characterized in
this model by linear $\gamma_{2}$ and nonlinear $\gamma_{3}$ rates. The
(assumed linear) inductive coupling between the FMSR and the loop antenna (LA)
[see Fig. \ref{FigSetup}(a)] is characterized by the rate $\gamma_{1}$.
Driving power (in units of rate) is denoted by $\Omega_{1}$, driving detuning angular frequency
is denoted by $\omega_{\mathrm{d}}$, and $\gamma=\gamma_{1}+\gamma_{2}$.
Stability analysis of Eq. (\ref{DKM E}), which has been performed in Ref.
\cite{Yurke_5054}, is summarized in SI section \ref{AppDKM}.

Bistability occurs in the region where the cubic polynomial equation
(\ref{DKM E}) has three real non--negative solutions. In the plane of
normalized driving detuning and power, this bistability region is red--colored
in the stability map shown in Fig. \ref{FigDK}(a). The normalized driving
detuning is given by $\omega_{\mathrm{dR}}=\omega_{\mathrm{d}}/\omega
_{\mathrm{dc}}$, and the normalized power by $\Omega_{1\mathrm{R}}=\Omega
_{1}/\Omega_{1\mathrm{c}}$. The boundary of the bistability region contains a
single cusp (bistability onset) point $\left(  \omega_{\mathrm{dc}}%
,\Omega_{1\mathrm{c}}\right)  $, labeled in Fig. \ref{FigDK}(a) by the symbol
$P_{\mathrm{c}}$. The values of $\omega_{\mathrm{dc}}$ and $\Omega
_{1\mathrm{c}}$ are given by Eqs. (\ref{omdc}) and (\ref{om1c}) in SI section
\ref{AppDKM}, respectively. Bistability occurs provided that
$\left\vert \omega_{\mathrm{K}}\right\vert \geq\sqrt{3}\gamma_{3}$. For the
plots in Fig. \ref{FigDK}(a-c), assumed parameters' values are listed in the
figure caption. The plot in Fig. \ref{FigDK}(c) displays the reflectivity
$\left\vert r\right\vert $ [see Eq. (\ref{DKM r}) in SI section \ref{AppDKM}] as a function of normalized detuning $\omega_{\mathrm{dR}}\ $for three
different values of normalized driving power $\Omega_{1\mathrm{R}}$, which are
labeled by the overlaid horizontal white dashed lines in Fig. \ref{FigDK}(a).
The normalized detuning at the peak point, which is denoted by $\omega
_{\mathrm{dpR}}$, is plotted in Fig. \ref{FigDK}(b). A flow map for the
Bosonic mode complex amplitude $C$ in the bistability region is presented in
Fig. \ref{FigDK}(d).

\textbf{FMSR} -- Both competing models predict nonlinear response and
multistability. In the current study, these effects are experimentally
explored using a FMSR made of yttrium iron garnet (YIG). As can be seen by
comparing the cubic equations (\ref{RDM cubic}) and (\ref{DKM E}), as well as
Figs. \ref{FigRD} and \ref{FigDK}, the two competing models yield clearly
distinguishable predictions, which are experimentally testable (see SI section
\ref{AppCCM}). Our experimental setup is designed to allow directly
testing models' predictions against measurements.

A sketch of the experimental setup is shown in Fig. \ref{FigSetup}(a). The
blue-colored radio frequency (RF) components allow both driving and detection
of FMSR magnetic resonance. The FMSR is inductively coupled to a microwave LA.
Magnetic resonances are identified using a vector network analyzer (VNA). A
radio frequency amplifier (RFA), and two phase-locked signal generators (SG)
are employed for driving, and the response is monitored using a radio
frequency spectrum analyzer (RFSA), which is serially connected to a 30 dB
attenuator. A circulator (C) and two splitter/combiner (SC) components are
employed to direct the input and output RF signals [see Fig.~\ref{FigSetup}(a)].

The angular frequency of the FMSR Kittel (uniform) mode $\omega_{\mathrm{m}}$
is approximately given by $\omega_{\mathrm{m}}=\mu_{0}\gamma_{\mathrm{e}%
}H_{\mathrm{s}}$ \cite{Walker_390}, where $\mathbf{H}_{\mathrm{s}}$ is the
static magnetic field, $H_{\mathrm{s}}=\left\vert \mathbf{H}_{\mathrm{s}%
}\right\vert $, $\mu_{0}$ is the free space permeability, and $\gamma
_{\mathrm{e}}/2\pi=28%
\operatorname{GHz}%
\operatorname{T}%
^{-1}$\ is the gyromagnetic ratio \cite{Fletcher_687}. The applied static
magnetic field $\mathbf{H}_{\mathrm{s}}$ is controlled by adjusting the
relative position of a magnetized Neodymium using a motorized stage [not shown
in the sketch in Fig. \ref{FigSetup}(a)].

The FMSR's response to externally--applied driving is probed using three
experimental methods. The first one, which is based on magneto--optical (MO)
modulation
\cite{Haigh_133602,Osada_103018,Sharma_087205,Liu_3698,Chai_820,Zhu_1291,Li_040344,Nayak_193905,Buks_486}%
, is demonstrated by the plot in Fig. \ref{FigSetup}(b). The Stokes and
anti-Stokes sidebands [labeled in Fig. \ref{FigSetup}(b) as S and A,
respectively] originate from mixing between simultaneously applied RF and
optical driving and photon--magnon Brillouin scattering
\cite{Ghasemian_12757,Liu_060405,Haigh_143601,Hisatomi_207401,Hisatomi_174427,Wu_023711}%
. The telecom band wavelength tunable laser (TL), polarization controller
(PC), single--mode optical fibers, and polarimeter optical spectrum analyzer
(POSA) \cite{Buks_486}, which are used for the MO modulation method, are
red--colored in the sketch shown in Fig. \ref{FigSetup}(a).

The plot in Fig. \ref{FigSetup}(c) demonstrates the second method, in which
the reflectivity $r$ is measured using a VNA connected to the LA [see Fig.
\ref{FigSetup}(a)]. The third method, which is discussed in SI section
\ref{AppIMD}, is based on intermodulation (IMD) of two
simultaneously--applied RF driving tones [see Fig. \ref{FigSetup}(d)]
\cite{Mathai_67001}. Bistability--induced hysteresis is demonstrated by the
plots in Fig. \ref{FigSetup}(e-f), which are performed using the MO modulation
method. For each driving power $P$, the driving frequency $f_{\mathrm{T}}$ in
Fig. \ref{FigSetup}(e) is swept upwards, whereas the frequency sweeping
direction is reversed for the plot shown in Fig. \ref{FigSetup}(f).

\begin{figure}[ptb]
\begin{center}
\includegraphics[width=3.0in,keepaspectratio]{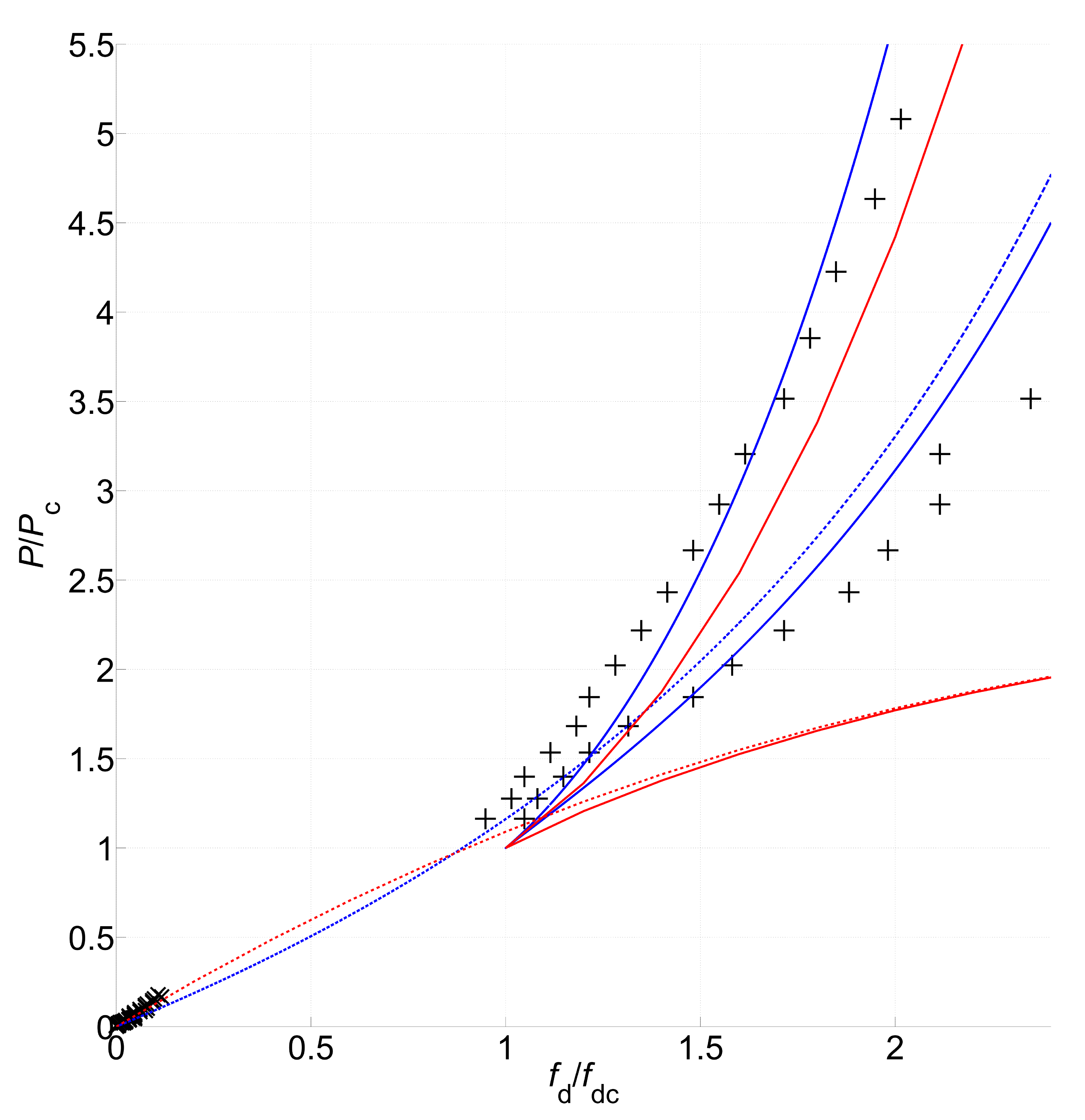}
\end{center}
\caption{{}Comparison with experimental results. Measured jump points are
labeled by the symbol $+$, whereas the symbol $\times$ is used for measured
peak points. The blue and red curves represent theoretical predictions based
on the RD and the Duffing--Kerr models, respectively. Calculated jump points
are represented by solid lines, whereas dashed lines label calculated peak
points. Jump points for the RD model are calculated using Eq. (\ref{RDM J de})
of SI section \ref{AppRDM}, whereas Eq. (\ref{DKM J omd}) of SI section \ref{AppDKM}
is employed for the Duffing--Kerr model. Optimized fit parameters' values are
$\gamma_{1}/\gamma=0.4$, $\omega_{\mathrm{K}}/\gamma=-0.01$ and $\gamma
_{3}=0.1\times3^{-1/2} \omega_{\mathrm{K}}$.}%
\label{FigSP}%
\end{figure}

\textbf{Comparison with experimental results} -- A comparison between the
measured FMSR response and predictions derived from both competing models is shown
in Fig. \ref{FigSP}. For relatively low driving power $P$, reflectivity
measurements are used to determine the peak frequency $f_{\mathrm{p}}$ [see
Fig. \ref{FigSetup}(c)]. Above a critical value of the driving power, which is
denoted by $P_{\mathrm{c}}$, bistability occurs in a region bounded between
two jump frequencies denoted by $f_{\mathrm{j}1}$ and $f_{\mathrm{j}2}$. Both
jump frequencies are probed using the MO method [see Fig. \ref{FigSetup}(b)].
The value of $f_{\mathrm{j}1}$ ($f_{\mathrm{j}2}$) is measured by sweeping the
driving frequency upwards (downwards) [see Fig. \ref{FigSetup}(e-f)]. In the
plot shown in Fig. \ref{FigSP}, each measured peak point having frequency
$f_{\mathrm{p}}$ and driving power $P_{\mathrm{p}}$ is represented by the
symbol $\times$ at the point $\left(  f_{\mathrm{d}}/f_{\mathrm{dc}%
},P_{\mathrm{p}}/P_{\mathrm{c}}\right)  $, where $f_{\mathrm{d}}%
=f_{\mathrm{p}}-f_{\mathrm{p0}}$, $f_{\mathrm{dc}}=f_{\mathrm{c}%
}-f_{\mathrm{p0}}$, $f_{\mathrm{p0}}$ is the peak frequency value in the limit
$P\rightarrow0$, and $f_{\mathrm{c}}$ is the peak frequency corresponding to the
bistability onset driving power, which is denoted by $P_{\mathrm{c}}$.
Measured jump points are represented by the symbol $+$. Theoretical
predictions based on the RD and the Duffing--Kerr models are represented by
blue and red curves, respectively. Calculated jump (peak) points are plotted
using solid (dashed) lines. Note that the stochastic nature of the jumping
process gives rise to relatively large scattering of the measured values of
the jump frequencies $f_{\mathrm{j}1}$ and $f_{\mathrm{j}2}$. The
data--theory comparison shown in Fig. \ref{FigSP} indicates that the
RD model better aligns with the experimental results.

\textbf{Discussion} -- The experimentally--observed bistability suggests that
the dynamics of the spin system under study is nonlinear. Nonlinearity is
introduced in the Duffing--Kerr model by implementing the method of
Bosonization. However, it has remained unclear how the Bosonization method,
which gives rise to bistability that is otherwise theoretically excluded,
can be justified
\cite{Katz_040404,Leppenen_2404_02134,Minganti_042118,Vicentini_013853,Landa_043601}%
. On the other hand, in the RD limit, the modified
master equation (\ref{MME}) yields predictions, which are found to be
consistent with the experimental results.

\textbf{Summary} -- Our findings indirectly support the hypothesis that
disentanglement spontaneously occurs in the spin system under study. The nonlinear term
added to the master equation (\ref{MME}) does not violate both norm
conservation and positivity of the density operator $\rho$ \cite{Buks_012439}.
The modified master equation (\ref{MME}) can be constructed for any physical
system whose Hilbert space has finite dimensionality. For a multipartite
system, disentanglement between any pair of subsystems can be introduced.
Disentanglement is invariant under any subsystem unitary transformation.
Disentanglement term has no effect on any product (i.e. disentangled) state,
thus, in the absence of entanglement, the added term does not vary any
prediction of standard QM. Disentanglement is applicable for both
distinguishable and indistinguishable particles \cite{Buks_2400248}. Further
study is needed to determine whether the spontaneous disentanglement
hypothesis is internally consistent, and whether its predictions are
consistent with experimental observations obtained with other physical systems.

\textbf{Acknowledgments} -- The author thank Lajos Diosi for intriguing
discussions. The research was supported by the Technion homeland
security foundation.


\bibliographystyle{ieeepes}
\bibliography{acompat,Eyal_Bib}

\widetext \widetext \clearpage

\begin{center}
\textbf{{\large Supporting information: Disentanglement--induced bistability
in a magnetic resonator}}

{\small {Eyal Buks} }

{\small {Andrew and Erna Viterbi Department of Electrical Engineering,
Technion, Haifa 32000 Israel} }
\end{center}

\setcounter{equation}{0} \setcounter{figure}{0} \setcounter{table}{0}
\setcounter{section}{0} \setcounter{page}{1}
\makeatletter \renewcommand{\theequation}{S\arabic{equation}}
\renewcommand{\thefigure}{S\arabic{figure}}
\renewcommand{\thesection}{S\arabic{section}}
\renewcommand{\bibnumfmt}[1]{[S#1]} \renewcommand{\citenumfont}[1]{S#1}

The supporting information is mainly devoted to the two competing theoretical models.

\section{Modified master equation}

\label{AppMME}

The modified master equation given by Eq. (\ref{MME}) in the main text is
based on Gorini-Kossakowski-Sudarshan-Lindblad linear master equation
\cite{SM_Fernengel_385701,SM_Lindblad_119,SM_Manzano_025106}, and on a
nonlinear extension giving rise to disentanglement.

\subsection{Lindblad superoperator}

The Lindblad superoperator $\mathcal{L}$ in Eq. (\ref{MME}) in the main text
is given by \cite{SM_carmichael2009open}%
\begin{align}
\mathcal{L}  &  =\sum_{l=1}^{L}\frac{\left(  \hat{n}_{0}+1\right)  \Gamma_{1}%
}{4}\mathcal{D}_{\rho}\left(  S_{l,-}\right)  +\frac{\hat{n}_{0}\Gamma_{1}}%
{4}\mathcal{D}_{\rho}\left(  S_{l,+}\right) \nonumber\\
&  +\frac{\left(  2\hat{n}_{0}+1\right)  \Gamma_{\varphi}}{2}\mathcal{D}%
_{\rho}\left(  S_{l,z}\right)  \;,\nonumber\\
&  \label{superoperator}%
\end{align}
where $l$ denotes spin index, and the Lindbladian $\mathcal{D}_{\rho}\left(
X\right)  $ for an operator $X$\ is given by%
\begin{equation}
\mathcal{D}_{\rho}\left(  X^{{}}\right)  =X^{{}}\rho X^{\dag}-\frac{X^{\dag
}X^{{}}\rho+\rho X^{\dag}X^{{}}}{2}\;. \label{Lindbladian}%
\end{equation}
The positive damping rates $\Gamma_{1}$ and $\Gamma_{\varphi}$, and the
thermal occupation factor $\hat{n}_{0}$, are related to the longitudinal
$T_{1}$ and the transverse $T_{2}$ relaxation times, and to the thermal
equilibrium spin polarization $P_{z0}$, by $1/T_{1}=\Gamma_{1}\left(  2\hat
{n}_{0}+1\right)  $, $1/T_{2}=\left(  \Gamma_{1}/2+\Gamma_{\varphi}\right)
\left(  2\hat{n}_{0}+1\right)  $ and $-1/P_{z0}=2\hat{n}_{0}+1$.

\subsection{Disentanglement}

The disentanglement term in the modified master equation given by Eq.
(\ref{MME}) in the main text is derived from a modified Schr\"{o}dinger
equation for the ket vector $\left\vert \psi\right\rangle $ having the form%
\begin{equation}
\frac{\mathrm{d}}{\mathrm{d}t}\left\vert \psi\right\rangle =\left[
-i\hbar^{-1}\mathcal{H}-\left(  \Theta-\left\langle \Theta\right\rangle
\right)  \right]  \left\vert \psi\right\rangle \;, \label{MSE}%
\end{equation}
where $\hbar$ is the Planck's constant, $\mathcal{H}^{{}}=\mathcal{H}^{\dag}$
is the Hamiltonian, the operator $\Theta$ is allowed to depend on $\left\vert
\psi\right\rangle $, and $\left\langle \Theta\right\rangle \equiv\left\langle
\psi\right\vert \Theta\left\vert \psi\right\rangle $. Formally, a master
equation for a density operator $\rho$ can be derived from a given
Schr\"{o}dinger equation for the time evolution of pure states. However, for
the case where the Schr\"{o}dinger equation is allowed to nonlinearly depend
on $\left\vert \psi\right\rangle $, strictly speaking, the obtained master
equation is valid only for pure states. In the current study, however, the
modified master equation given by Eq. (\ref{MME}) in the main text is treated
as applicable for a general mixed state.

The operator $\Theta$ [see Eq. (\ref{MSE})] is given by $\Theta=\gamma
_{\mathrm{D}}\mathcal{Q}^{\left(  \mathrm{D}\right)  }$, where the rate
$\gamma_{\mathrm{D}}$ is positive, and the operator $\mathcal{Q}^{\left(
\mathrm{D}\right)  }$\ is Hermitian. The construction of the operator
$\mathcal{Q}^{\left(  \mathrm{D}\right)  }$ is explained below for a general
multipartite system composed of three subsystems labeled as 'a', 'b' and 'c'
\cite{SM_Buks_012439}. The Hilbert space of the system $H=H_{\mathrm{a}%
}\otimes H_{\mathrm{b}}\otimes H_{\mathrm{c}}$ is a tensor product of
subsystem Hilbert spaces $H_{\mathrm{a}}$, $H_{\mathrm{b}}$ and $H_{\mathrm{c}%
}$. The dimensionality of the Hilbert space $H_{\mathrm{L}}$ of subsystem
$\mathrm{L}$, which is denoted by $d_{\mathrm{L}}$, where $\mathrm{L}%
\in\left\{  \mathrm{a},\mathrm{b},\mathrm{c}\right\}  $, is assumed to be
finite. A general observable of subsystem $\mathrm{L}$ can be expanded using
the set of generalized Gell-Mann matrices $\left\{  \lambda_{1}^{\left(
\mathrm{L} \right)  },\lambda_{2}^{\left(  \mathrm{L}\right)  },\cdots
,\lambda_{d_{\mathrm{L}}^{2}-1}^{\left(  \mathrm{L}\right)  }\right\}  $.
Entanglement between subsystems a and b can be quantified
\cite{SM_Schlienz_4396,SM_Peres_1413,SM_Hill_5022,SM_Wootters_1717,SM_Coffman_052306,SM_Vedral_2275,SM_Eltschka_424005,SM_Dur_062314,SM_Coiteux_200401,SM_Takou_011004,SM_Elben_200501}
by the nonnegative variable $\tau_{\mathrm{ab}}$, which is given by
$\tau_{\mathrm{ab}}=\left\langle \mathcal{Q}_{\mathrm{ab}}^{\left(
\mathrm{D}\right)  }\right\rangle $, where the operator $\mathcal{Q}%
_{\mathrm{ab}}^{\left(  \mathrm{D}\right)  }$ is given by%
\begin{equation}
\mathcal{Q}_{\mathrm{ab}}^{\left(  \mathrm{D}\right)  }=\eta_{\mathrm{ab}%
}\operatorname{Tr}\left(  C^{\mathrm{T}}\left\langle C\right\rangle \right)
\;, \label{Q_12 Tr}%
\end{equation}
and where $\eta_{\mathrm{ab}}$ is a positive constant. The $\left(
a,b\right)  $ entry of the $\left(  d_{\mathrm{a}}^{2}-1\right)  \times\left(
d_{\mathrm{b}}^{2}-1\right)  $ matrix $C$ is the observable $\mathcal{C}%
\left(  \lambda_{a}^{\left(  \mathrm{a}\right)  },\lambda_{b}^{\left(
\mathrm{b}\right)  }\right)  $, and the $\left(  a,b\right)  $ entry of the
$\left(  d_{\mathrm{a}}^{2}-1\right)  \times\left(  d_{\mathrm{b}}%
^{2}-1\right)  $ matrix $\left\langle C\right\rangle $ is its expectation
value $\left\langle \mathcal{C}\left(  \lambda_{a}^{\left(  \mathrm{a}\right)
},\lambda_{b}^{\left(  \mathrm{b}\right)  }\right)  \right\rangle $. For any
given observable $O_{\mathrm{a}}^{{}}=O_{\mathrm{a}}^{\dag}$ of subsystem a,
and a given observable $O_{\mathrm{b}}^{{}}=O_{\mathrm{b}}^{\dag}$ of
subsystem b, the observable $\mathcal{C}\left(  O_{\mathrm{a}},O_{\mathrm{b}%
}\right)  $ is defined by $\mathcal{C}\left(  O_{\mathrm{a}},O_{\mathrm{b}%
}\right)  =O_{\mathrm{a}}\otimes O_{\mathrm{b}}\otimes I_{\mathrm{c}%
}-\left\langle O_{\mathrm{a}}\otimes I_{\mathrm{b}}\otimes I_{\mathrm{c}%
}\right\rangle \left\langle I_{\mathrm{a}}\otimes O_{\mathrm{b}}\otimes
I_{\mathrm{c}}\right\rangle $, where $I_{\mathrm{L}}$ is the $d_{\mathrm{L}%
}\times d_{\mathrm{L}}$ identity matrix, and where $\mathrm{L}\in\left\{
\mathrm{a},\mathrm{b},\mathrm{c}\right\}  $. The entanglement variable
$\tau_{\mathrm{ab}}$ is invariant under any single subsystem unitary
transformation. In a similar way, a disentanglement operator $\mathcal{Q}%
^{\left(  \mathrm{D}\right)  }$ corresponding to any given pair of subsystems
can be defined.

\section{Driven $L$ spin 1/2}

\label{AppDLS}

The driven $L$ spin 1/2 Hamiltonian $\mathcal{H}$ is given by $\hbar
^{-1}\mathcal{H}=-\left(  \omega_{0}/2\right)  S_{z}+\left(  \omega
_{\mathrm{K}}/8\right)  \left(  S_{+}S_{-}+S_{-}S_{+}\right)  +\left(
\omega_{\mathrm{A}}/8\right)  \left(  S_{+}^{2}+S_{-}^{2}\right)  +\left(
\omega_{1}/4\right)  \left(  S_{+}e^{i\omega_{\mathrm{T}}t}+S_{-}%
e^{-i\omega_{\mathrm{T}}t}\right)  $ [see Eq. (\ref{H DLS}) in the main text].
The following commutation relations hold $\left[  S_{i},S_{j}\right]
=2i\epsilon_{ijk}S_{k}$, $\left[  S_{z},S_{\pm}\right]  =\pm2S_{\pm}$, and
$\left[  S_{+},S_{-}\right]  =4S_{z}$ (recall that $S_{\pm}=S_{x}\pm iS_{y})$.
The Heisenberg equations of motion for the operators $S_{+}$ and $S_{z}$\ are
given by%
\begin{align}
\frac{\mathrm{d}S_{+}}{\mathrm{d}t}  &  =-i\omega_{0}S_{+}-\frac
{i\omega_{\mathrm{K}}\left(  S_{+}S_{z}+S_{z}S_{+}\right)  }{2}\nonumber\\
&  -\frac{i\omega_{\mathrm{A}}\left(  S_{-}S_{z}+S_{z}S_{-}\right)  }%
{2}-i\omega_{1}e^{-i\omega_{\mathrm{T}}t}S_{z}\;,\nonumber\\
&  \label{d S_+ / dt}%
\end{align}
and%
\begin{equation}
\frac{\mathrm{d}S_{z}}{\mathrm{d}t}=\frac{i\omega_{\mathrm{A}}\left(
S_{-}^{2}-S_{+}^{2}\right)  }{2}+\frac{i\omega_{1}\left(  e^{-i\omega
_{\mathrm{T}}t}S_{-}-e^{i\omega_{\mathrm{T}}t}S_{+}\right)  }{2}\;.
\label{d S_z / dt H}%
\end{equation}
In terms of the rotating operators $S_{\mathrm{R}\pm}=e^{\pm i\omega
_{\mathrm{T}}t}S_{\pm}$ the equations of motion are expressed as%
\begin{align}
\frac{\mathrm{d}S_{\mathrm{R}+}}{\mathrm{d}t}  &  =i\omega_{\mathrm{d}%
}S_{\mathrm{R}+}-\frac{i\omega_{\mathrm{K}}\left(  S_{\mathrm{R}+}S_{z}%
+S_{z}S_{\mathrm{R}+}\right)  }{2}\nonumber\\
&  -i\omega_{1}S_{z}-\frac{i\omega_{\mathrm{A}}\left(  S_{\mathrm{R}-}%
S_{z}+S_{z}S_{\mathrm{R}-}\right)  e^{2i\omega_{\mathrm{T}}t}}{2}%
\;,\nonumber\\
&  \label{d Sigma_+ / dt}%
\end{align}
where $\omega_{\mathrm{d}}=\omega_{\mathrm{T}}-\omega_{0}$, and%
\begin{equation}
\frac{\mathrm{d}S_{z}}{\mathrm{d}t}=\frac{i\omega_{1}\left(  S_{\mathrm{R}%
-}-S_{\mathrm{R}+}\right)  }{2}+\frac{i\omega_{\mathrm{A}}\left(
S_{\mathrm{R}-}^{2}e^{2i\omega_{\mathrm{T}}t}-S_{\mathrm{R}+}^{2}%
e^{-2i\omega_{\mathrm{T}}t}\right)  }{2}\;. \label{d S_z / dt}%
\end{equation}

\section{Driven two spin 1/2}

\label{AppDTS}

The matrix representation of the vector operator $\mathbf{S}=\mathbf{S}%
_{\mathrm{a}}+\mathbf{S}_{\mathrm{b}}=\left(  S_{x},S_{y},S_{z}\right)  $,
which represents the total angular momentum vector in units of $\hbar/2$, is
given by (the first and second spins are labelled by the letters a and b,
respectively)%
\begin{align}
S_{x}  &  =S_{\mathrm{a}x}+S_{\mathrm{b}x}\nonumber\\
&  \dot{=}\left(
\begin{array}
[c]{cccc}%
0 & 0 & 1 & 0\\
0 & 0 & 0 & 1\\
1 & 0 & 0 & 0\\
0 & 1 & 0 & 0
\end{array}
\right)  +\left(
\begin{array}
[c]{cccc}%
0 & 1 & 0 & 0\\
1 & 0 & 0 & 0\\
0 & 0 & 0 & 1\\
0 & 0 & 1 & 0
\end{array}
\right) \nonumber\\
&  =\left(
\begin{array}
[c]{cccc}%
0 & 1 & 1 & 0\\
1 & 0 & 0 & 1\\
1 & 0 & 0 & 1\\
0 & 1 & 1 & 0
\end{array}
\right)  \;,\nonumber\\
&
\end{align}%
\begin{align}
S_{y}  &  =S_{\mathrm{a}y}+S_{\mathrm{b}y}\nonumber\\
&  \dot{=}\left(
\begin{array}
[c]{cccc}%
0 & 0 & -i & 0\\
0 & 0 & 0 & -i\\
i & 0 & 0 & 0\\
0 & i & 0 & 0
\end{array}
\right)  +\left(
\begin{array}
[c]{cccc}%
0 & -i & 0 & 0\\
i & 0 & 0 & 0\\
0 & 0 & 0 & -i\\
0 & 0 & i & 0
\end{array}
\right) \nonumber\\
&  =\left(
\begin{array}
[c]{cccc}%
0 & -i & -i & 0\\
i & 0 & 0 & -i\\
i & 0 & 0 & -i\\
0 & i & i & 0
\end{array}
\right)  \;,\nonumber\\
&
\end{align}
and%
\begin{align}
S_{z}  &  =S_{\mathrm{a}z}+S_{\mathrm{b}z}\nonumber\\
&  \dot{=}\left(
\begin{array}
[c]{cccc}%
1 & 0 & 0 & 0\\
0 & 1 & 0 & 0\\
0 & 0 & -1 & 0\\
0 & 0 & 0 & -1
\end{array}
\right)  +\left(
\begin{array}
[c]{cccc}%
1 & 0 & 0 & 0\\
0 & -1 & 0 & 0\\
0 & 0 & 1 & 0\\
0 & 0 & 0 & -1
\end{array}
\right) \nonumber\\
&  =\left(
\begin{array}
[c]{cccc}%
2 & 0 & 0 & 0\\
0 & 0 & 0 & 0\\
0 & 0 & 0 & 0\\
0 & 0 & 0 & -2
\end{array}
\right)  \;.\nonumber\\
&
\end{align}
Note that $\left[  S_{i},S_{j}\right]  =2i\epsilon_{ijk}S_{k}$. The operators
$S_{\pm}$ are defined by $S_{\pm}=S_{x}\pm iS_{y}$, and the following holds%
\begin{align}
S_{+}  &  \dot{=}\left(
\begin{array}
[c]{cccc}%
0 & 2 & 2 & 0\\
0 & 0 & 0 & 2\\
0 & 0 & 0 & 2\\
0 & 0 & 0 & 0
\end{array}
\right)  \;,\\
S_{-}  &  \dot{=}\left(
\begin{array}
[c]{cccc}%
0 & 0 & 0 & 0\\
2 & 0 & 0 & 0\\
2 & 0 & 0 & 0\\
0 & 2 & 2 & 0
\end{array}
\right)  \;,
\end{align}
$\left[  S_{z},S_{\pm}\right]  =\pm2S_{\pm}$ and $\left[  S_{+},S_{-}\right]
=4S_{z}$.

The Hamiltonian is given by $\mathcal{H}=\mathcal{H}_{0}+\mathcal{H}%
_{\mathrm{p}}$. The static part $\mathcal{H}_{0}$ is given by%
\begin{equation}
\frac{\mathcal{H}_{0}}{\hbar}=\frac{\omega_{0}S_{z}}{2}+\frac{\omega
_{\mathrm{K}}\left(  S_{+}S_{-}+S_{-}S_{+}\right)  }{8}\;, \label{H0 2SD}%
\end{equation}
where $\omega_{0}$ and $\omega_{\mathrm{K}}$ are real constants, where
\begin{equation}
\frac{S_{+}S_{-}+S_{-}S_{+}}{8}\dot{=}\left(
\begin{array}
[c]{cccc}%
1 & 0 & 0 & 0\\
0 & 1 & 1 & 0\\
0 & 1 & 1 & 0\\
0 & 0 & 0 & 1
\end{array}
\right)  \;,
\end{equation}
thus%
\begin{equation}
\frac{\mathcal{H}_{0}}{\hbar}\dot{=}\left(
\begin{array}
[c]{cccc}%
\omega_{\mathrm{K}}+\omega_{0} & 0 & 0 & 0\\
0 & \omega_{\mathrm{K}} & \omega_{\mathrm{K}} & 0\\
0 & \omega_{\mathrm{K}} & \omega_{\mathrm{K}} & 0\\
0 & 0 & 0 & \omega_{\mathrm{K}}-\omega_{0}%
\end{array}
\right)  \;. \label{H0 M 2SD}%
\end{equation}
The driving term $\mathcal{H}_{\mathrm{p}}$ of the Hamiltonian is given by%
\begin{equation}
\frac{\mathcal{H}_{\mathrm{p}}}{\hbar}=\frac{\omega_{1}\left(  S_{+}
e^{-i\omega_{\mathrm{T}}t}+S_{-}e^{i\omega_{\mathrm{T}}t}\right)  }{4}\;,
\end{equation}
where the transverse driving power $\omega_{1}$ and angular frequency
$\omega_{\mathrm{T}}$ are real constants.

A rotating frame transformation yields%
\begin{equation}
-iu_{0}^{\dag}\frac{\mathrm{d}u_{0}^{{}}}{\mathrm{d}t}+u_{0}^{\dag}%
\frac{\mathcal{H}_{{}}}{\hbar}u_{0}^{{}}=\omega_{\mathrm{K}}I+\frac
{\mathcal{H}_{\mathrm{R}}}{\hbar}\;,
\end{equation}
where%
\begin{equation}
u_{0}\dot{=}\left(
\begin{array}
[c]{cccc}%
e^{-i\omega_{\mathrm{T}}t} & 0 & 0 & 0\\
0 & 1 & 0 & 0\\
0 & 0 & 1 & 0\\
0 & 0 & 0 & e^{i\omega_{\mathrm{T}}t}%
\end{array}
\right)  \;,
\end{equation}
$I$ is the identity operator, and the matrix representation of $\mathcal{H}%
_{\mathrm{R}}$ is given by%
\begin{equation}
\frac{\mathcal{H}_{\mathrm{R}}}{\hbar}\dot{=}\left(
\begin{array}
[c]{cccc}%
-\omega_{\mathrm{d}} & \frac{\omega_{1}}{2} & \frac{\omega_{1}}{2} & 0\\
\frac{\omega_{1}}{2} & 0 & \omega_{\mathrm{K}} & \frac{\omega_{1}}{2}\\
\frac{\omega_{1}}{2} & \omega_{\mathrm{K}} & 0 & \frac{\omega_{1}}{2}\\
0 & \frac{\omega_{1}}{2} & \frac{\omega_{1}}{2} & \omega_{\mathrm{d}}%
\end{array}
\right)  \;,
\end{equation}
where $\omega_{\mathrm{d}}=\omega_{\mathrm{T}}-\omega_{0}$ is the angular
detuning frequency.

\section{Rapid disentanglement model}

\label{AppRDM}

The term $S_{\mathrm{R}+}S_{z}+S_{z}S_{\mathrm{R}+}$ in Eq.
(\ref{d Sigma_+ / dt}), which originates from dipolar coupling, can be
expressed as $S_{\mathrm{R}+}S_{z}+S_{z}S_{\mathrm{R}+}=\sum_{l^{\prime
},l^{\prime\prime}=1}^{L}\left(  S_{\mathrm{R}l^{\prime},+}S_{l^{\prime\prime
},z}+S_{z,l^{\prime}}S_{\mathrm{R}l^{\prime\prime},+}\right)  $. The
approximation $\left\langle S_{\mathrm{R}l^{\prime},+}S_{l^{\prime\prime}
,z}+S_{z,l^{\prime}}S_{\mathrm{R}l^{\prime\prime},+}\right\rangle
\simeq\left\langle S_{\mathrm{R}l^{\prime},+}\right\rangle \left\langle
S_{l^{\prime\prime},z}\right\rangle +\left\langle S_{z,l^{\prime}%
}\right\rangle \left\langle S_{\mathrm{R}l^{\prime\prime},+}\right\rangle
\simeq2L^{-2}\left\langle S_{\mathrm{R}+}\right\rangle \left\langle
S_{z}\right\rangle $ can be implemented provided that the rate of
disentanglement $\gamma_{\mathrm{D}}$\ is sufficiently large. Damping is taken
into account by adding decay terms to Eqs. (\ref{d Sigma_+ / dt}) and
(\ref{d S_z / dt}) [see Eq. (\ref{superoperator})], which become [recall that
the term proportional to $\omega_{\mathrm{A}}$ in the Hamiltonian
(\ref{H DLS}) is disregarded in the rotating wave approximation (RWA)]%
\begin{equation}
\frac{\mathrm{d}P_{+}}{\mathrm{d}t}=i\omega_{\mathrm{d}}P_{+}-i\omega
_{\mathrm{K}}P_{+}P_{z}-i\omega_{1}P_{z}-\frac{P_{+}}{T_{2}}\;,
\label{d P_+ / dt}%
\end{equation}
and%
\begin{equation}
\frac{\mathrm{d}P_{z}}{\mathrm{d}t}=\frac{i\omega_{1}\left(  P_{-}
-P_{+}\right)  }{2}-\frac{P_{z}-P_{z0}}{T_{1}}\;, \label{d P_z / dt}%
\end{equation}
where $P_{\pm}=\left\langle S_{\mathrm{R}\pm}\right\rangle $, $P_{z}%
=\left\langle S_{z}\right\rangle $, $P_{z0}$ represents the steady state value
of $P_{z}$ for the case where $\omega_{1}=0$ (no driving), and $T_{1}^{-1}$
($T_{2}^{-1}$) is the longitudinal (transverse) damping, rate.

In steady state, i.e. for $\mathrm{d}P_{+}/\mathrm{d}t=0$ and $\mathrm{d}%
P_{z}/\mathrm{d}t=0$, Eqs. (\ref{d P_+ / dt}) and (\ref{d P_z / dt}) yield%
\begin{equation}
P_{z}=\frac{1+\left(  \omega_{\mathrm{d}}-\omega_{\mathrm{K}}P_{z}\right)
^{2}T_{2}^{2}}{1+\left(  \omega_{\mathrm{d}}-\omega_{\mathrm{K}}P_{z}\right)
^{2}T_{2}^{2}+\omega_{1}^{2}T_{1}T_{2}}P_{z0}\;, \label{eq for P_z}%
\end{equation}
and%
\begin{equation}
P_{+}=\frac{i\omega_{1}T_{2}P_{z}}{i\left(  \omega_{\mathrm{d}}-\omega
_{\mathrm{K}}P_{z}\right)  T_{2}-1}\;.
\end{equation}
The relation (\ref{eq for P_z}) can be expressed as $F\left(  P_{z}%
/P_{z0},\omega_{\mathrm{d}}T_{2}\right)  =0$, where the function $F$ is given
by%
\begin{equation}
F\left(  z,\delta\right)  =z\left(  1+\left(  \delta-4\sqrt{D}z\right)
^{2}+2W\right)  -1-\left(  \delta-4\sqrt{D}z\right)  ^{2}\;, \label{F(z,de)=}%
\end{equation}
where $D=\left(  \omega_{\mathrm{K}}T_{2}P_{z0}/4\right)  ^{2}$ and
$W=\omega_{1}^{2}T_{1}T_{2}/2$.

\subsection{Peak points}

The cubic polynomial equation (\ref{eq for P_z}) for $P_{z}$ yields%
\begin{equation}
\frac{\mathrm{d}P_{z}}{\mathrm{d}\omega_{\mathrm{d}}}=-\frac{2T_{2}^{2}\left(
P_{z}-P_{z0}\right)  \left(  \omega_{\mathrm{d}}-\omega_{\mathrm{K}}
P_{z}\right)  }{\mathcal{D}}\;. \label{d P_z / d omega_d}%
\end{equation}
where $\mathcal{D}=T_{2}^{2}\left(  \omega_{\mathrm{d}}-\omega_{\mathrm{K}%
}P_{z}\right)  \left(  \omega_{\mathrm{d}}+2\omega_{\mathrm{K}}P_{z0}%
-3\omega_{\mathrm{K}}P_{z}\right)  +1+\omega_{1}^{2}T_{1}T_{2}$. Thus, at peak
points, for which $0=\partial P_{z}/\partial\omega_{\mathrm{d}}$, one has
$\omega_{\mathrm{d}}=\omega_{\mathrm{K}}P_{z}$. This relation together with
the condition $F=0$ yield%
\begin{equation}
\delta=\frac{4\sqrt{D}}{1+2W}\;. \label{RDM delta PP}%
\end{equation}

\subsection{Bistability onset points}

At a bistability onset point the following three conditions hold
\begin{equation}
0=\frac{\mathrm{d}\delta}{\mathrm{d}z}=-\frac{F_{z}}{F_{\delta}}\;,\label{C1}%
\end{equation}%
\begin{equation}
0=\frac{\mathrm{d}^{2}\delta}{\mathrm{d}z^{2}}=-\frac{F_{\delta}^{2}%
F_{zz}-2F_{z}F_{\delta}F_{z\delta}+F_{z}^{2}F_{\delta\delta}}{F_{\delta}^{3}%
}\;,\label{C2}%
\end{equation}
and
\begin{equation}
0=F\left(  z,\delta\right)  \;,\label{C3}%
\end{equation}
where $F$ with an added subscript denotes a partial derivative, e.g.
$F_{z}=\partial F/\partial z$. The first condition (\ref{C1}), which can be
expressed as $0=F_{z}=1+\delta^{2}-16\delta\sqrt{D}z+48Dz^{2}+2W+8\sqrt
{D}\delta-32Dz$, implies that the second condition (\ref{C2}) can be expressed
as $0=F_{zz}=96Dz-32D-16\sqrt{D}\delta$. By extracting the value of $W$ from
the condition $0=F_{z}$ (\ref{C1}), and the value of $\delta$ from the
condition $0=F_{zz}$ (\ref{C2}), the condition $0=F\left(  z,\delta\right)  $
(\ref{C3}) can be expressed as a cubic polynomial equation for $z$, whose
solutions are given by $z_{1}=Z\left(  q\right)  $, and $z_{\pm}=Z\left(
qe^{\pm\frac{2\pi i}{3}}\right)  $, where the function $Z$ is defined by
$Z\left(  \zeta\right)  =\left(  \zeta+1/\zeta+3\right)  /4$, and where%
\begin{equation}
q=\sqrt[3]{\frac{\left(  1+\sqrt{1-D}\right)  ^{2}}{D}}\;.\label{q DTS}%
\end{equation}
For a given solution for $z$, the corresponding dimensionless detuning is
$\delta=2\sqrt{D}\left(  3z-1\right)  $ [see condition (\ref{C2})], and
dimensionless driving power is $W=\left(  -1-\delta^{2}+16\delta\sqrt
{D}z-48Dz^{2}-8\sqrt{D}\delta+32Dz\right)  /2=6D\left(  1-z\right)  ^{2}-1/2$
[see condition (\ref{C1})].

Solutions for $z$ are acceptable provided that $\operatorname{Im}z=0$ and
$0\leq z\leq1$. These conditions are satisfied by the solutions $z_{+}$ and
$z_{-}$\ in the range $D\geq1$ (for $D<1$ bistability is excluded). The
detuning and driving power corresponding to $z_{\pm}$ are denoted by
$\delta_{\pm}$ and $W_{\pm}$, respectively. Note that in the range $D\geq1$
[see Eq. (\ref{q DTS})]%
\begin{equation}
q=e^{\frac{2i\tan^{-1}\left(  \sinh x\right)  }{3}}\;,
\end{equation}
where $D=\cosh^{2}x$, and the following holds%
\begin{equation}
z_{\pm}=\frac{3+2\cos\left(  \frac{2\left(  \tan^{-1}\left(  \sinh x\right)
\pm\pi\right)  }{3}\right)  }{4}\;.\label{z pm}%
\end{equation}
For $D=1$ (i.e. for $x=0$ and $q=1$) the following holds $z_{\pm}=1/2$,
$\delta=1$ [recall that $\delta=2\sqrt{D}\left(  3z-1\right)  $], and $W=1$
[recall that $W=6D\left(  1-z\right)  ^{2}-1/2$]. Note that $z_{+}%
\rightarrow1/4$ and $z_{-}\rightarrow1$ in the limit $D\rightarrow\infty$. The
asymptotic expansion $\tan^{-1}\left(  \sinh x\right)  \simeq\pi/2-1/\sinh x$,
which is valid in the limit $x\gg1$, yields $z_{-}\simeq1-\left(  \sqrt
{3}/6\right)  /\sqrt{D-1}$ [see Eq. (\ref{z pm}), and recall that $D=\cosh
^{2}x$ and that $\cosh^{2}x-1=\sinh^{2}x$], and thus for this limit
$W_{-}\simeq1/\left(  2D\right)  $, and%
\begin{equation}
\frac{W_{+}}{W_{-}}\simeq\frac{27D^{2}}{4}\;.\label{W_+ / W_-}%
\end{equation}

\subsection{Jump points}

For any $W\in\left(  W_{-},W_{+}\right)  $, bistability region is bounded
between two normalized detuning frequencies denoted by $\delta_{\mathrm{j}1}$
and $\delta_{\mathrm{j}2}$, which are found by solving conditions (\ref{C1})
and (\ref{C3}), which can be expressed as%
\begin{equation}
0=z\left(  1+\left(  \delta-4\sqrt{D}z\right)  ^{2}+2W\right)  -1-\left(
\delta-4\sqrt{D}z\right)  ^{2}\;, \label{RDM JR1}%
\end{equation}
and%
\begin{equation}
0=1+2W+\left(  \delta-4\sqrt{D}z\right)  \left(  \delta-4\sqrt{D}\left(
3z-2\right)  \right)  \;. \label{RDM JR2}%
\end{equation}
By solving Eq. (\ref{RDM JR2}) for $W$, and substituting the solution into Eq.
(\ref{RDM JR1}), one finds that%
\begin{equation}
\delta=4\sqrt{D}z^{2}\pm\sqrt{16z^{2}\left(  z-1\right)  ^{2}D-1}\;.
\label{RDM J de}%
\end{equation}
Note that $16z^{2}\left(  z-1\right)  ^{2}\in\left[  0,1\right]  $ for
$z\in\left[  0,1\right]  $.

\section{Nutation}

\label{AppNut}

In this section quantum (Heisenberg) and classical (Poisson) equations of
motion are derived from the Hamiltonian (\ref{H DLS}) in the main text, and
terms proportional to $\omega_{\mathrm{A}}$, which are disregarded in the RWA,
and which can give rise to nutation, are kept.

The Heisenberg equation of motion (\ref{d S_+ / dt}) can be rewritten as%
\begin{align*}
&  \frac{\mathrm{d}}{\mathrm{d}t}\left(
\begin{array}
[c]{c}%
S_{+}\\
S_{-}%
\end{array}
\right) \\
&  =\left(
\begin{array}
[c]{cc}%
-i\omega_{0} & 0\\
0 & i\omega_{0}%
\end{array}
\right)  \left(
\begin{array}
[c]{c}%
S_{+}\\
S_{-}%
\end{array}
\right) \\
&  +\left(
\begin{array}
[c]{cc}%
-\frac{i\omega_{\mathrm{K}}}{2} & -\frac{i\omega_{\mathrm{A}}}{2}\\
\frac{i\omega_{\mathrm{A}}}{2} & \frac{i\omega_{\mathrm{K}}}{2}%
\end{array}
\right)  \left(  \left(
\begin{array}
[c]{c}%
S_{+}\\
S_{-}%
\end{array}
\right)  S_{z}+S_{z}\left(
\begin{array}
[c]{c}%
S_{+}\\
S_{-}%
\end{array}
\right)  \right) \\
&  -i\omega_{1}\left(
\begin{array}
[c]{c}%
e^{-i\omega_{\mathrm{T}}t}\\
-e^{i\omega_{\mathrm{T}}t}%
\end{array}
\right)  S_{z}\;.\\
&
\end{align*}
The transformation%
\begin{equation}
\left(
\begin{array}
[c]{c}%
\mathcal{S}_{+}\\
\mathcal{S}_{-}%
\end{array}
\right)  =\left(
\begin{array}
[c]{cc}%
e^{i\omega_{\mathrm{T}}t} & 0\\
0 & e^{-i\omega_{\mathrm{T}}t}%
\end{array}
\right)  \left(
\begin{array}
[c]{cc}%
X & Y\\
Y & X
\end{array}
\right)  \left(
\begin{array}
[c]{c}%
S_{+}\\
S_{-}%
\end{array}
\right)  \;,
\end{equation}
where $X=\left(  1/2\right)  \left(  \sqrt{1+\omega_{\mathrm{A}}%
/\omega_{\mathrm{K}}}+\sqrt{1-\omega_{\mathrm{A}}/\omega_{\mathrm{K}}}\right)
$ and where $Y=\left(  1/2\right)  \left(  \sqrt{1+\omega_{\mathrm{A}}%
/\omega_{\mathrm{K}}}-\sqrt{1-\omega_{\mathrm{A}}/\omega_{\mathrm{K}}}\right)
$, yields [note that $X^{2}+Y^{2}=1$, $2XY=\omega_{\mathrm{A}}/\omega
_{\mathrm{K}}$, and $X^{2}-Y^{2}=\sqrt{1-\left(  \omega_{\mathrm{A}}%
/\omega_{\mathrm{K}}\right)  ^{2}}$]%
\begin{align}
&  \frac{\mathrm{d}}{\mathrm{d}t}\left(
\begin{array}
[c]{c}%
\mathcal{S}_{+}\\
\mathcal{S}_{-}%
\end{array}
\right) \nonumber\\
&  =i\left(
\begin{array}
[c]{cc}%
\omega_{\mathrm{T}}-\omega_{0}^{\prime} & \omega_{0}^{\prime}\frac
{\omega_{\mathrm{A}}}{\omega_{\mathrm{K}}}e^{2i\omega_{\mathrm{T}}t}\\
-\omega_{0}^{\prime}\frac{\omega_{\mathrm{A}}}{\omega_{\mathrm{K}}%
}e^{-2i\omega_{\mathrm{T}}t} & -\omega_{\mathrm{T}}+\omega_{0}^{\prime}%
\end{array}
\right)  \left(
\begin{array}
[c]{c}%
\mathcal{S}_{+}\\
\mathcal{S}_{-}%
\end{array}
\right) \nonumber\\
&  +\frac{i\omega_{0}\omega_{\mathrm{K}}}{2\omega_{0}^{\prime}}\left(
\begin{array}
[c]{c}%
-\mathcal{S}_{+}S_{z}-S_{z}\mathcal{S}_{+}\\
\mathcal{S}_{-}S_{z}+S_{z}\mathcal{S}_{-}%
\end{array}
\right) \nonumber\\
&  -i\omega_{1}\left(
\begin{array}
[c]{c}%
X-Ye^{2i\omega_{\mathrm{T}}t}\\
-X+Ye^{-2i\omega_{\mathrm{T}}t}%
\end{array}
\right)  S_{z}\;,\nonumber\\
&
\end{align}
where $\omega_{0}^{\prime}=\omega_{0}/\sqrt{1-\left(  \omega_{\mathrm{A}%
}/\omega_{\mathrm{K}}\right)  ^{2}}$.

Alternatively, real equations of motion are derived below. In terms of $S_{x}$
and $S_{y}$ [recall that $S_{\pm}=S_{x}\pm iS_{y}$, $S_{+}S_{-}+S_{-}%
S_{+}=2\left(  S_{x}^{2}+S_{y}^{2}\right)  $ and $S_{+}^{2}+S_{-}^{2}=2\left(
S_{x}^{2}-S_{y}^{2}\right)  $] the Hamiltonian (\ref{H DLS}) in the main text
can be expressed as%
\begin{equation}
\frac{\mathcal{H}}{\hbar}=-\frac{\mathbf{S}\cdot\mathbf{S}_{\mathrm{H}}}{2}\;,
\end{equation}
where $\mathbf{S}_{\mathrm{H}}=\mathbf{S}_{0}-\left(  \omega_{\mathrm{K}%
}/2\right)  \mathbf{S}_{\mathrm{D}}+\mathbf{S}_{1}$, $\mathbf{S}_{0}%
=\omega_{0}\mathbf{\hat{z}}$, $\mathbf{S}_{\mathrm{D}}=\left(  1+\omega
_{\mathrm{A}}/\omega_{\mathrm{K}}\right)  S_{x}\mathbf{\hat{x}}+\left(
1-\omega_{\mathrm{A}}/\omega_{\mathrm{K}}\right)  S_{y}\mathbf{\hat{y}}$,
$\mathbf{S}_{1}=-\omega_{1}\left(  \cos\left(  \omega_{\mathrm{T}}t\right)
\mathbf{\hat{x}}-\sin\left(  \omega_{\mathrm{T}}t\right)  \mathbf{\hat{y}%
}\right)  $, and the equations of motion can be expressed as (recall that
$\left[  S_{i},S_{j}\right]  =2i\epsilon_{ijk}S_{k}$)%
\begin{equation}
\frac{\mathrm{d}\mathbf{S}}{\mathrm{d}t}=\frac{\mathbf{S}\times\mathbf{S}%
_{\mathrm{H}}-\mathbf{S}_{\mathrm{H}}\times\mathbf{S}}{2}\;.
\end{equation}
Classical equation of motion is obtained by treating $\mathbf{S}$ and
$\mathbf{S}_{\mathrm{H}}$\ as real 3D vectors ($\mathbf{S}$ is replaced by
$\mathbf{P}$)%
\begin{equation}
\frac{\mathrm{d}\mathbf{P}}{\mathrm{d}t}=\mathbf{P}\times\mathbf{P}%
_{\mathrm{H}}-\mathbf{P}_{\mathrm{T}}\;,
\end{equation}
where the added damping term $\mathbf{P}_{\mathrm{T}}$ is given by
$\mathbf{P}_{\mathrm{T}}=\left(  P_{x}/T_{2},P_{y}/T_{2},\left(  P_{z}%
-S_{z0}\right)  /T_{1}\right)  $.

\section{Intermodulation conversion gain}

\label{AppIMD}

In this section, the intermodulation conversion gain $g_{\mathrm{IMD}}$\ is
evaluated to lowest nonvanishing order in $\omega_{\mathrm{A}}$. Expressing
$S_{\mathrm{R}\pm}$ and $S_{z}$ as $S_{\mathrm{R}\pm}=P_{\pm}+p_{\pm}$ and
$S_{z}=P_{z}+z_{+}e^{2i\omega_{\mathrm{T}}t}+z_{-}e^{-2i\omega_{\mathrm{T}}t}%
$, neglecting rapidly rotating terms (at angular frequencies $\pm
\omega_{\mathrm{T}}$\ and $\pm2\omega_{\mathrm{T}}$), implementing mean field
approximation (MFA), treating $P_{\pm}$ and $P_{z}$ as constants and treating
$p_{\pm}$ and $z_{+}$\ as small, and adding damping terms yield [see Eqs.
(\ref{d Sigma_+ / dt}) and (\ref{d S_z / dt})]%
\begin{equation}
0=\left(  i\omega_{\mathrm{dK}}-\frac{1}{T_{2}}\right)  P_{+}-i\omega_{1}%
P_{z}\;, \label{P_+ ss}%
\end{equation}%
\begin{align}
\frac{\mathrm{d}p_{+}}{\mathrm{d}t}  &  =\left(  i\omega_{\mathrm{dK}}%
-\frac{1}{T_{2}}\right)  p_{+}\nonumber\\
&  -\frac{i\omega_{\mathrm{A}}\left(  \left(  P_{-}+p_{-}\right)  z_{-}%
+z_{-}\left(  P_{-}+p_{-}\right)  \right)  }{2}\;,\nonumber\\
&  \label{d p_+ / dt}%
\end{align}%
\begin{equation}
0=\frac{i\omega_{1}\left(  P_{-}-P_{+}\right)  }{2}-\frac{P_{z}-P_{z0}}{T_{1}%
}\;, \label{P_z ss}%
\end{equation}
and%
\begin{align}
\frac{\mathrm{d}z_{+}}{\mathrm{d}t}+\left(  2i\omega_{\mathrm{T}}+\frac
{1}{T_{1}}\right)  z_{+}  &  =\frac{i\omega_{\mathrm{A}}\left(  P_{-}%
+p_{-}\right)  ^{2}}{2}\;,\label{d z_+ / dt}\\
\frac{\mathrm{d}z_{-}}{\mathrm{d}t}+\left(  -2i\omega_{\mathrm{T}}+\frac
{1}{T_{1}}\right)  z_{-}  &  =-\frac{i\omega_{\mathrm{A}}\left(  P_{+}%
+p_{+}\right)  ^{2}}{2}\;, \label{d z_- / dt}%
\end{align}
where $\omega_{\mathrm{dK}}=\omega_{\mathrm{d}}-\omega_{\mathrm{K}}P_{z}$.
Note that $\omega_{\mathrm{dK}}$ vanishes at peak points, and that steady
state values of $P_{\pm}$ and $P_{z}$ are independent of $\omega_{\mathrm{A}}$
[see Eqs. (\ref{P_+ ss}) and (\ref{P_z ss})]. The steady state solution of
Eqs. (\ref{d z_+ / dt}) and (\ref{d z_- / dt}), which is given by%
\begin{equation}
z_{\pm}=\frac{i\omega_{\mathrm{A}}\left(  P_{\mp}+p_{\mp}\right)  ^{2}%
}{2\left(  2i\omega_{\mathrm{T}}\pm\frac{1}{T_{1}}\right)  }\;,
\end{equation}
yields [see Eq. (\ref{d p_+ / dt}), recall that $p_{\pm}$, $z_{+}$\ and
$\omega_{\mathrm{A}}$\ are treated as small, and note that zeroth order terms
are disregarded]%
\begin{equation}
\frac{\mathrm{d}p_{+}}{\mathrm{d}t}=\left(  i\omega_{\mathrm{dK}}-\frac
{1}{T_{2}}\right)  p_{+}+\frac{\omega_{\mathrm{A}}^{2}P_{+}^{2}p_{-}}{2\left(
2i\omega_{\mathrm{T}}-\frac{1}{T_{1}}\right)  }\;,
\end{equation}
or in a matrix form%
\begin{equation}
\frac{\mathrm{d}}{\mathrm{d}t}\left(
\begin{array}
[c]{c}%
p_{+}\\
p_{-}%
\end{array}
\right)  =\left(
\begin{array}
[c]{cc}%
W_{1}^{{}} & W_{2}^{{}}\\
W_{2}^{\ast} & W_{1}^{\ast}%
\end{array}
\right)  \left(
\begin{array}
[c]{c}%
p_{+}\\
p_{-}%
\end{array}
\right)  =W\left(
\begin{array}
[c]{c}%
p_{+}\\
p_{-}%
\end{array}
\right)  \;,
\end{equation}
where $W_{1}=i\omega_{\mathrm{dK}}-1/T_{2}$ and $W_{2}=\left(  \omega
_{\mathrm{A}}^{2}/2\right)  P_{+}^{2}/\left(  2i\omega_{\mathrm{T}}%
-1/T_{1}\right)  $. The trace $T_{\mathrm{W}}=W_{1}+W_{1}^{\ast}$ and the
determinant $D_{\mathrm{W}}=\left\vert W_{1}\right\vert ^{2}-\left\vert
W_{2}\right\vert ^{2}$ of the $2\times2$ matrix $W$\ are given by
$T_{\mathrm{W}}=-2/T_{2}$ and $D_{\mathrm{W}}=\omega_{\mathrm{dK}}^{2}%
+1/T_{2}^{2}-\left(  \omega_{\mathrm{A}}^{4}/4\right)  \left\vert
P_{+}\right\vert ^{4}/\left(  4\omega_{\mathrm{T}}^{2}+1/T_{1}^{2}\right)  $.
The intermodulation conversion gain $g_{\mathrm{IMD}}$ at angular frequency
$\omega$ is given by [see Eq. (83) of Ref. \cite{SM_Yurke_5054}]%
\begin{equation}
g_{\mathrm{IMD}}=\left\vert \frac{2\gamma_{1}W_{2}}{\left(  \lambda
_{+}-i\omega\right)  \left(  \lambda_{-}-i\omega\right)  }\right\vert ^{2}\;,
\label{g_IMD}%
\end{equation}
where $\lambda_{\pm}=T_{\mathrm{W}}/2\pm\sqrt{\left(  T_{\mathrm{W}}/2\right)
^{2}-D_{\mathrm{W}}}$ are the eigenvalues of the matrix $W$.

\section{Duffing--Kerr model}

\label{AppDKM}

In the Holstein--Primakoff transformation \cite{SM_Holstein_1098}, the
operators $S_{\pm}=S_{x}\pm iS_{y}$ and $S_{z}$ are expressed as
$S_{+}=2B^{\dag}\left(  L-B^{\dag}B^{{}}\right)  ^{1/2}$, $S_{-}=2\left(
L-B^{\dag}B^{{}}\right)  ^{1/2}B^{{}}$ and $S_{z}=-L+2B^{\dag}B^{{}}$, where
$L$ is the total number of spins, and where $B^{\dag}B$ is a number operator.
Consistency with the commutation relations $\left[  S_{z},S_{\pm
}\right]  =\pm2S_{\pm}$ and $\left[  S_{+},S_{-}\right]  =4S_{z}$ is obtained
by assuming that the operators $B^{{}}$ and $B^{\dag}$\ satisfy the Bosonic
commutation relation $\left[  B,B^{\dag}\right]  =1$.

The term proportional to $\omega_{1}$ in the Hamiltonian $\mathcal{H}$ given
by Eq. (\ref{H DLS}) in the main text represents transverse driving. In the
Bosonization method, this term is excluded from the closed--system Hamiltonian
$\mathcal{H}$, and transverse driving is instead accounted for by introducing
an external feedline \cite{SM_Gardiner_3761}. The coupling between the
feedline and the spins, which is assumed to be linear, is characterized by a
rate denoted by $\gamma_{1}$. By employing the relation%
\begin{equation}
\frac{S_{+}S_{-}+S_{-}S_{+}}{8}=\left(  L-1\right)  B^{\dag}B^{{}}-B^{\dag
}B^{\dag}B^{{}}B^{{}}+\frac{L}{2}\;,
\end{equation}
and the approximation $\left(  L-B^{\dag}B^{{}}\right)  ^{1/2}\simeq L^{1/2}$
one finds that the closed-system Hamiltonian $\mathcal{H}$ [see Eq.
(\ref{H DLS}) in the main text] can be replaced by the Hamiltonian
$\mathcal{H}_{\mathrm{HP}}$, which is given by (recall that in the RWA the
term proportional to $\omega_{\mathrm{A}}$\ is disregarded, and note that all
constant terms in $\mathcal{H}_{\mathrm{HP}}$ are disregarded, and it is
assumed that $L\gg1$)
\cite{SM_Hill_S227,SM_Wang_224410,SM_Zhang_987511,SM_Mathai_67001}%
\begin{equation}
\frac{\mathcal{H}_{\mathrm{HP}}}{\hbar}=-\tilde{\omega}_{0}B^{\dag}B^{{}%
}-\omega_{\mathrm{K}}B^{\dag}B^{\dag}B^{{}}B^{{}}\;,\label{H_HP}%
\end{equation}
where $\tilde{\omega}_{0}=\omega_{0}-L\omega_{\mathrm{K}}$.

The results given below are based on Ref. \cite{SM_Yurke_5054}, which reports
on a theoretical study of a system having an Hamiltonian similar to
$\mathcal{H}_{\mathrm{HP}}$. Linear and nonlinear damping are characterized by
the rates $\gamma_{2}$ and $\gamma_{3}$, respectively. The rate $\gamma_{1}$ depends on the
coupling between the FMSR and the LA. The MFA yields in steady state $\left\langle B\right\rangle =C$, where the
complex number $C$ is found by solving%
\begin{equation}
C=-\frac{i\sqrt{2\gamma_{1}\Omega_{1}}}{\left(  -i\omega_{\mathrm{d}}%
+\gamma\right)  +\left(  i\omega_{\mathrm{K}}+\gamma_{3}\right)  \left\vert
C\right\vert ^{2}}\;,\label{eq for B}%
\end{equation}
$\Omega_{1}$ is the transverse driving power, $\omega_{\mathrm{d}}$ is the
driving detuning angular frequency, and $\gamma=\gamma_{1}+\gamma_{2}$. The
condition for the complex amplitude $C$ (\ref{eq for B}) yields a cubic
polynomial equation for the real non--negative variable $E=\left\vert
C\right\vert ^{2}$ given by%
\begin{equation}
E^{3}+\frac{2\left(  \gamma\gamma_{3}-\omega_{\mathrm{d}}\omega_{\mathrm{K}%
}\right)  E^{2}}{\omega_{\mathrm{K}}^{2}+\gamma_{3}^{2}}+\frac{\left(
\omega_{\mathrm{d}}^{2}+\gamma^{2}\right)  E}{\omega_{\mathrm{K}}^{2}%
+\gamma_{3}^{2}}-\frac{2\gamma_{1}\Omega_{1}}{\omega_{\mathrm{K}}^{2}%
+\gamma_{3}^{2}}=0\;.\label{eq for E}%
\end{equation}
The reflection amplitude $r$ off the LA, which is inductively coupled to the
FMSR, is given by%
\begin{equation}
r=\frac{-i\omega_{\mathrm{d}}-\gamma_{-}+\left(  i\omega_{\mathrm{K}}%
+\gamma_{3}\right)  E}{-i\omega_{\mathrm{d}}+\gamma+\left(  i\omega
_{\mathrm{K}}+\gamma_{3}\right)  E}\;,\label{DKM r}%
\end{equation}
where $\gamma_{-}=\gamma_{1}-\gamma_{2}$. Expressions for peak, bistability
onset and jump points (defined in section \ref{AppRDM} above) are derived below.

\subsection{Peak points}

The cubic equation (\ref{eq for E}) yields a peak point occurring at
$\omega_{\mathrm{d}}=\omega_{\mathrm{K}}E$. This condition together with Eq.
(\ref{eq for E}) yield (recall that $E\geq0$)%
\begin{equation}
\omega_{\mathrm{d}}=\omega_{\mathrm{K}}\frac{2\gamma_{1}\Omega_{1}}{\gamma
^{2}}+O\left(  \Omega_{1}^{2}\right)  \;. \label{DKM omd PP}%
\end{equation}

\subsection{The bistability onset point}

No bistability onset points occur when $\left\vert \omega_{\mathrm{K}%
}\right\vert <\sqrt{3}\gamma_{3}$, and a single bistability onset point occurs
when $\left\vert \omega_{\mathrm{K}}\right\vert \geq\sqrt{3}\gamma_{3}$. At
that point the driving power is given by $\Omega_{1}=\Omega_{1\mathrm{c}}$,
the angular detuning frequency by $\omega_{\mathrm{d}}=\omega_{\mathrm{dc}}$,
and the photon number by $E=E_{\mathrm{c}}$, where \cite{SM_Yurke_5054}%
\begin{align}
\Omega_{1\mathrm{c}}  &  =\frac{\left(  \omega_{\mathrm{K}}^{2}+\gamma_{3}
^{2}\right)  E_{\mathrm{c}}^{3}}{2\gamma_{1}}\;,\label{om1c}\\
\omega_{\mathrm{dc}}  &  =\gamma\frac{\omega_{\mathrm{K}}}{\left\vert
\omega_{\mathrm{K}}\right\vert }\frac{4\gamma_{3}|\omega_{\mathrm{K}}
|+\sqrt{3}\left(  \omega_{\mathrm{K}}^{2}+\gamma_{3}^{2}\right)  }%
{\omega_{\mathrm{K}}^{2}-3\gamma_{3}^{2}}\;,\label{omdc}\\
E_{\mathrm{c}}  &  =\frac{2\gamma}{\sqrt{3}\left(  \left\vert \omega
_{\mathrm{K}}\right\vert -\sqrt{3}\gamma_{3}\right)  }\;. \label{Ec}%
\end{align}

\subsection{Jump points}

At jump points the following holds (the derivation is similar to the one
employed above in section \ref{AppRDM})%
\begin{equation}
\omega_{\mathrm{d}}=2\omega_{\mathrm{K}}E\pm\sqrt{\left(  \left(
\omega_{\mathrm{K}}^{2}-3\gamma_{3}^{2}\right)  E-4\gamma\gamma_{3}\right)
E-\gamma^{2}}\;. \label{DKM J omd}%
\end{equation}

\section{Comparison between the two competing models}

\label{AppCCM}

As can be seen from the stability map shown in Fig. (\ref{FigDK})(a) in the
main text, in the Duffing--Kerr model bistability is possible in a driving
power $\Omega_{1}$ range that is only bounded from below $\Omega_{1}%
>\Omega_{1\mathrm{c}}$ [see Eq. (\ref{om1c}) of section \ref{AppDKM}]. In
contrast, for the RD model, the bistability driving power $W$ range has both
lower $W_{-}$ and upper $W_{+}$ bounds [see Fig. \ref{FigRD}(a) in the main
text]. Moreover, in the RD model, bistability is possible provided that
$D=\left(  \omega_{\mathrm{K}}T_{2}P_{z0}/4\right)  ^{2}\geq1$, whereas in the
Duffing--Kerr model, bistability is possible provided that $\left(
1/3\right)  \left(  \omega_{\mathrm{K}}/\gamma_{3}\right)  ^{2}\geq1$.

The FMSR Stoner--Wohlfarth energy $E_{\mathrm{M}}$ is given by $E_{\mathrm{M}%
}/V_{\mathrm{s}}=-\mu_{0}\mathbf{M}\cdot\mathbf{H}_{\mathrm{s}}+K_{\mathrm{c1}%
}\sin^{2}\phi+K_{\mathrm{c2}}\sin^{4}\phi$, where $\mathbf{M}$ is the
magnetization vector, $K_{\mathrm{c1}}$ ($K_{\mathrm{c2}}$) is the first
(second) order anisotropy constant, $V_{\mathrm{s}}=4\pi R_{\mathrm{s}}^{3}/3$
is the volume of the FMSR having radius $R_{\mathrm{s}}$, and $\phi$ is the
angle between $\mathbf{M}$ and the the easy axis \cite{SM_Blunde_Mag}. It is
assumed that the FMSR is fully magnetized, i.e. $\left\vert \mathbf{M}%
\right\vert \simeq M_{\mathrm{s}}$, where $M_{\mathrm{s}}$ is the saturation
magnetization. For the case where $\mathbf{H}_{\mathrm{s}}$ is applied
parallel to the FMSR easy axis, the rate $\omega_{\mathrm{K}}$ in Eq.
(\ref{H DLS}) in the main text is given by $\omega_{\mathrm{K}}=2\gamma
_{\mathrm{e}}K_{\mathrm{c1}}/\left(  N_{\mathrm{s}}M_{\mathrm{s}}\right)  $,
where $N_{\mathrm{s}}=V_{\mathrm{s}}\rho_{\mathrm{s}}$ is the total number of
FMSR spins, and $\rho_{\mathrm{s}}$ is the spin density. For YIG
$M_{\mathrm{s}}=140%
\operatorname{kA}%
/%
\operatorname{m}%
$, $K_{\mathrm{c1}}=-610%
\operatorname{J}%
/%
\operatorname{m}%
^{3}$, and $\rho_{\mathrm{s}}=4.2\times10^{21}%
\operatorname{cm}%
^{-3}$ \cite{SM_Stancil_Spin}, thus $\omega_{\mathrm{K}}=-7.\,1\times10^{-9}%
\operatorname{Hz}%
$\ for a sphere of radius $R_{\mathrm{s}}=125%
\operatorname{\mu m}%
$.

The approximation $W_{+}/W_{-}\simeq27D^{2}/4$, which is valid in the limit
$D\gg1$ [see Eq. (\ref{W_+ / W_-}) of section \ref{AppRDM}], implies that for
our setup $W_{+}$ [i.e. the bistability upper bound for $W$, see Fig.
\ref{FigRD}(a) in the main text] is far too high to be experimentally
accessible. For the same limit of $D\gg1$\ the approximation $W_{-}%
\simeq1/\left(  2D\right)  $ implies that at the low bistability onset point
$\omega_{1}\simeq1/\sqrt{T_{1}T_{2}D}=4T_{1}^{-1/2}T_{2}^{-3/2}/\left\vert
\omega_{\mathrm{K}}\right\vert $ [recall that $D=\left(  \omega_{\mathrm{K}%
}T_{2}P_{z0}/4\right)  ^{2}$, and that it is assumed that the FMSR is fully
magnetized]. For comparison, for the Duffing--Kerr model, in the absence of
nonlinear damping, i.e. for $\gamma_{3}=0$, at the bistability onset point
$\Omega_{1}=\left(  4/3^{3/2}\right)  \left(  \gamma^{3}/\gamma_{1}\right)
/\left\vert \omega_{\mathrm{K}}\right\vert $ [see Eqs. (\ref{om1c}) and
(\ref{Ec}) of section \ref{AppDKM}].

\end{document}